\documentclass[fleqn,usenatbib]{mnras}
\usepackage{graphicx,amssymb,natbib,cite,aas_macros,amsmath}
\usepackage{txfonts}
\usepackage{graphics,times,placeins,float}
\usepackage{longtable,lscape,rotating}
\usepackage{supertabular,caption,multicol}
\usepackage[T1]{fontenc}
\usepackage{ae,aecompl}

\title[The mass-loss history of SN 2014C]{The peculiar mass-loss history of SN 2014C as revealed through AMI radio observations}
\author[G. E. Anderson et al.]{G. E. Anderson,$^{1,2}$\thanks{E-mail: gemma.anderson@curtin.edu.au} 
A. Horesh,$^{3,4}$ 
K. P. Mooley,$^2$ 
A. P. Rushton,$^{2,5}$ 
R. P. Fender,$^2$
\newauthor 
T. D. Staley,$^2$
M. K. Argo,$^{6,7}$
R. J. Beswick,$^{7}$
P. J. Hancock,$^{1,8}$
M. A. P\'erez-Torres,$^{9,10}$
\newauthor
Y. C. Perrott,$^{11}$ 
R. M. Plotkin,$^{1}$
M. L. Pretorius,$^{2}$
C. Rumsey,$^{11}$  
and D. J. Titterington$^{11}$ 	\\
$^1$International Centre for Radio Astronomy Research, Curtin University, GPO Box U1987, Perth, WA 6845, Australia\\
$^2$Department of Physics, Astrophysics, University of Oxford, Denys Wilkinson Building, Oxford, OX1 3RH, UK\\
$^3$Racah Institute of Physics, Hebrew University, Jerusalem, 91904, Israel\\
$^4$Benoziyo Center for Astrophysics, Weizmann Institute of Science, 76100 Rehovot, Israel\\
$^5$School of Physics and Astronomy, University of Southampton, Highfield, Southampton SO17 1BJ, UK\\
$^6$Jeremiah Horrocks Institute, University of Central Lancashire, Preston PR1 2HE, UK\\
$^7$Jodrell Bank Centre for Astrophysics, School of Physics and Astronomy, The University of Manchester,\\ ~Oxford Road, Manchester M13 9PL, UK\\ 
$^8$ARC Centre of Excellence for All-sky Astrophysics (CAASTRO)\\
$^9$Instituto de Astrofísica de Andalucia (IAA-CSIC), E-18008 Granada, Spain\\
$^{10}$Departamento de Fisica Teorica, Facultad de Ciencias, Universidad de Zaragoza, Spain\\
$^{11}$Astrophysics Group, Cavendish Laboratory, 19 J J Thomson Avenue, Cambridge CB3 0HE, UK\\
}

\date{Accepted 2016 December 15. Received 2016 November 8; in original form 2016 July 28}

\pagerange{\pageref{firstpage}--\pageref{lastpage}} \pubyear{2016}

\begin{document}

\maketitle

\label{firstpage}

\begin{abstract}

We present a radio light curve of supernova (SN) 2014C taken with the Arcminute Microkelvin Imager (AMI) Large Array at 15.7 GHz. Optical observations presented by Milisavljevic et al. demonstrated that SN 2014C metamorphosed from a stripped-envelope Type Ib SN into a strongly interacting Type IIn SN within 1 year. The AMI light curve clearly shows two distinct radio peaks, the second being a factor of 4 times more luminous than the first peak. This double bump morphology indicates two distinct phases of mass-loss from the progenitor star with the transition between density regimes occurring at 100-200 days. This reinforces the interpretation that SN 2014C exploded in a low density region before encountering a dense Hydrogen-rich shell of circumstellar material that was likely ejected by the progenitor prior to the explosion. The AMI flux measurements of the first light curve bump are the only reported observations taken within $\sim50$ to $\sim125$ days post-explosion, before the blast-wave encountered the Hydrogen shell. Simplistic synchrotron self-absorption (SSA) and free-free absorption (FFA) modelling suggest that some physical properties of SN 2014C, such as the mass-loss rate, are consistent with the properties of other Type Ibc and IIn SNe. However, our single frequency data does not allow us to distinguish between these two models, which implies they are likely too simplistic to describe the complex environment surrounding this event. Lastly, we present the precise radio location of SN 2014C obtained with eMERLIN, which will be useful for future VLBI observations of the SN.

\end{abstract}


\section{Introduction}

Mass-loss is an important ingredient in understanding the evolution of massive stars and has a significant impact on their final stellar death. The different types of core-collapse supernovae (SNe) we observe are a direct result of the amount of mass shed by the progenitor star. The majority of core-collapse SNe are Type II, which exhibit Hydrogen (H) emission lines in their optical spectra and in most cases are likely the last evolutionary stage of red supergiants \citep[RSG;][]{smith14}. Core-collapse Type Ib and Ic SNe (SN Ibc) show no H (and in the latter case no Helium [He]) in their optical spectra, which indicates that the progenitors went through a phase of extreme mass-loss before their inevitable demise, which stripped the majority (if not all) of their outer H (and He) envelopes. As a result, SNe Ibc are often referred to as ``stripped-envelope" SNe.

The commonly accepted paradigm when modelling stellar evolution is that massive stars evolve in isolation and that mass-loss is caused by line-driven winds that are highly dependent on metallicity \citep{conti76}. However, recent studies have shown that the influence of line driven-winds is likely overestimated and that other mechanisms may also play a substantial role in stellar mass-loss, particularly with the stripping of a massive star's H-envelope \citep[for example see][and references therein]{smith11,langer12,smith14}. These mechanisms include binary interactions involving Roche-lobe overflow, which may play a significant role in creating most SNe Ibc progenitors \citep[e.g.][]{iben85,yoon10,smith11,dessart12,bersten12,bersten14,yoon15}. During this binary evolution, a Hydrogen-rich (H-rich) common envelope can also be ejected \citep{podsiadlowski92}. 

Another form of extreme mass loss can occur when the progenitor star undergoes a very rapid and massive ejection of stellar material shortly before its SN. Such events are known as SN precursors or ``impostors" \citep{vandyk00,vandyk12} and are often part of the luminous blue variable (LBV) evolutionary phase where the ejected mass of material can be anywhere from $0.1$~M$_{\odot}$ up to $\geq10$~M$_{\odot}$ \citep{smith06}. LBVs have also been shown to be the progenitors of Type IIn SNe (SN IIn), which have strong narrow H lines \citep{gal-yam07,gal-yam09}. The most famous case of a SN precursor is 2009ip, which finally went SN in 2012 \citep{smith14mauerhan}. Studies by \citet{ofek14} demonstrate that the progenitors of more than $50\%$ of SNe IIn undergo at least one massive pre-explosion outburst within a year of their SN explosion. The narrow H lines seen in the optical spectra of SNe IIn are therefore caused by the SN shock encountering a dense Hydrogen shell (H-shell) of circumstellar material (CSM) that was lost by the progenitor up to many years before core collapse \citep[e.g.][]{chugai94,smith11,chugai03,chugai04,chugai06,kiewe12}. 

The interaction between the SN shock wave and the surrounding CSM gives rise to non-thermal radio and X-ray radiation. The radio light curve shows a typical turn on and power-law rise as higher frequencies become less subject to absorption processes, followed  by a power-law decline as the source become optically thin \citep{chevalier98}. The detection of radio emission allows us to probe the circumstellar environment of the SN and ultimately study the mass-loss history of the progenitor, particularly in its final phases of evolution. However, there are several examples of stripped-envelope supernova (SNe Ibc) that diverge from the canonical light curve behaviour \citep[e.g.][]{wellons12} and in some extreme cases show very late-time radio rebrightening 
\citep[e.g. SN 2001em, SN 2003bg, SN 2003gk, SN 2008iz, and PTF 11qcj,][]{schinzel09,salas13,bietenholz14,corsi14,kimani16}. Such events indicate a more extreme stellar environment than could be provided by constant mass loss from the progenitor's stellar wind, rather implying previous phases of more extreme mass loss and/or binary evolution.

Here we present the radio light curve of SN 2014C, an extremely unusual stripped-envelope supernova located in the nearby galaxy NGC 7331 ($z=0.0027$), that also shows late-time radio rebrightening. SN 2014C was discovered on 2014 Jan 5.1 UT with the Katzman Automatic Imaging Telescope (KAIT) at Lick Observatory, with the earliest optical detection in a pre-discovery image obtained on 2014 Jan 2.1 \citep{zheng14,kim14}. Early spectroscopic follow-up identified it as a SN Ib \citep{tartaglia14} and it was quickly detected at radio wavelengths with the VLA and CARMA \citep{kamble14,zauderer14}. However, spectroscopic observations obtained between 100 and 500 days post-explosion demonstrated that SN 2014C underwent a slow metamorphosis into a SN IIn within $\sim1$ year as indicated by the strengthening of H$\alpha$ emission combined with a very strong increase in radio and X-ray emission \citep{milisavljevic15,margutti16pp}. These authors interpret this spectral and temporal evolution as the result of the SN exploding in a low density region. The SN ejecta then encountered a dense H-rich CSM of $\sim1 \mathrm{M}_{\odot}$ that was likely ejected by the progenitor star decades or centuries before the SN event. It is likely that the ejecta of the SN first encountered this CSM shell sometime between Feb and May 2014, around 20 to 130 days after the explosion, when SN 2014C was a daylight object and therefore unobservable by optical telescopes. Based on these findings \citet{milisavljevic15} and \citet{margutti16pp} suggest that the H-envelope may have been ejected due to binary interactions that resulted in a common envelope evolution, an eruptive LBV event, or instabilities during the final stages of nuclear burning. SN 2014C also showed late-time mid-infrared (IR) rebrightening at $\sim250$ days resulting from the pre-existing dust in the surrounding environment \citep{tinyanont16pp}. However, these authors favour a LBV-like eruption scenario as the derived dust mass and radius properties are consistent with such an event occurring $\sim100$ years before the SN. 

\begin{figure*}
\begin{center}
\includegraphics[width=0.8\textwidth]{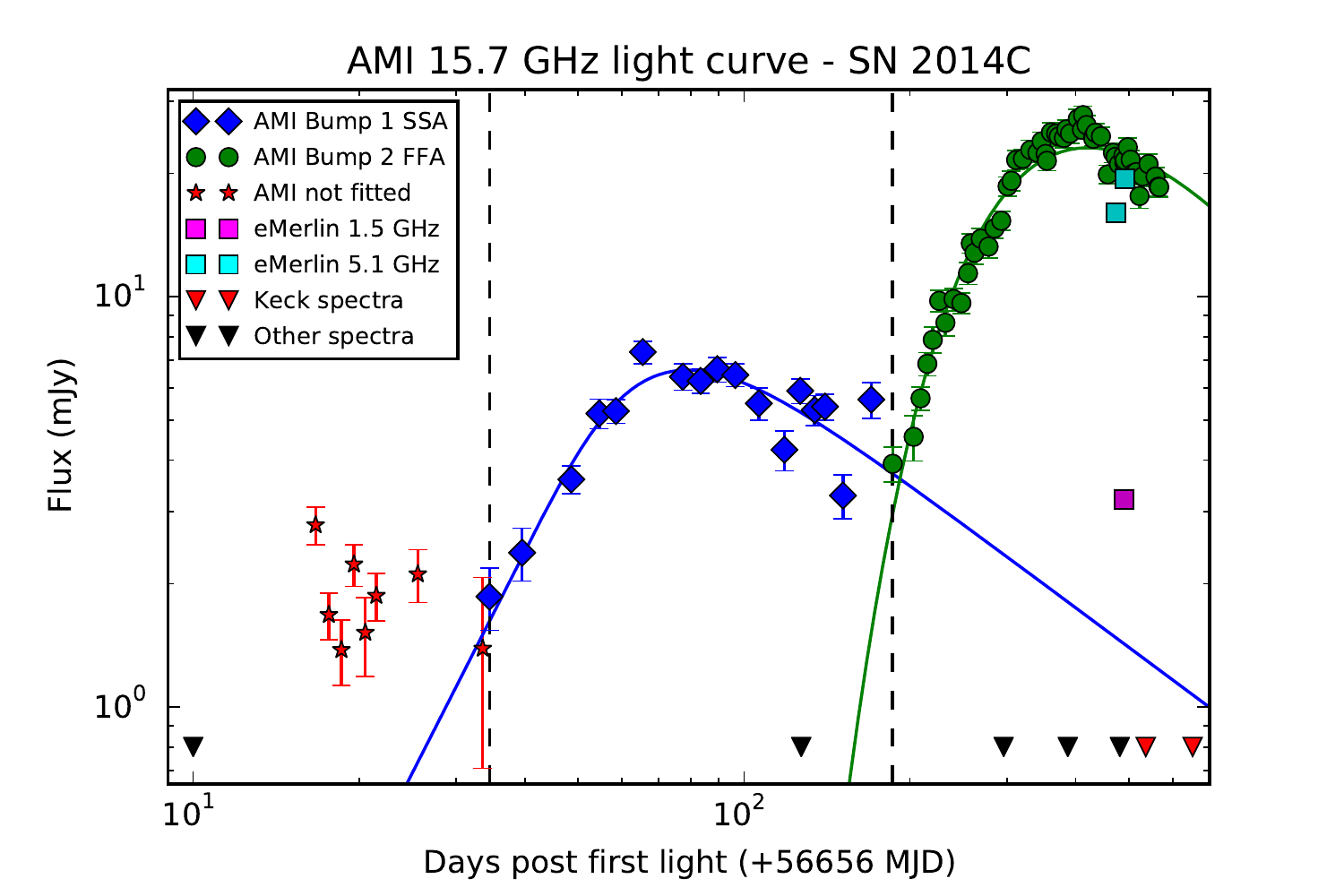}
\end{center}
\caption{The AMI 15.7 GHz light curve of SN 2014C. SN data points that make up bump 1 (blue diamonds) and bump 2 (green circles) were modelled separately. The best fit SSA and FFA model (see Table~\ref{tab:model}) are shown in the same colours for bumps 1 and 2, respectively. The vertical dashed lines indicate the start times we defined for bumps 1 and 2 (34.5 and 186.2 days post-burst) for modelling purposes. Data points not included in the modelling (red stars) are early time AMI detections of SN 2014C. The 1.5 and 5.1 GHz eMERLIN detections of SN 2014C are also included on the light curve. The observation times of our Keck spectra and the optical spectra described in \citep{milisavljevic15} are denoted by red and black triangles, respectively.}\label{fig:1}
\end{figure*}

In this paper we present the extremely high cadence 15.7 GHz radio light curve of SN 2014C taken with the Arcminute Microkelvin Imager (AMI) Large Array, which supports the findings of the shock encountering a high density CSM shell at $\sim100-200$ days post explosion as suggested by \citet{milisavljevic15},  \citet{margutti16pp} and \citet{tinyanont16pp}. Not only do we detect a late-time rise in radio emission as described by \citet{margutti16pp}, but also an early time bump in the radio light curve that occurs within $\sim100$ days of the explosions, demonstrating that the SN shock wave encountered a lower density environment at early times before encountering the H-shell (see Figure~\ref{fig:1}). The details and results of our AMI Large Array observations are presented in Section 2, along with new high resolution radio observations obtained with eMERLIN and optical spectra obtained with Keck. We provide a detailed description of the phases of the radio evolution of SN 2014C as seen with AMI Large Array in Section 3. Section 4 describes the standard radio modelling we performed to obtain the physical properties of the SN shock wave and CSM. We then discuss these physical properties with respect to SNe Ibc, SNe IIn, and those obtained by other authors for SN 2014C in Section 5. In this analysis we assume that first light was Dec 30, 2013 $\pm 1$ day (MJD 56656 $\pm 1$), which was derived by \citet{margutti16pp} through modelling the bolometric light curve, and we use the Cepheid distance of $14.7 \pm 0.6$ Mpc to the host galaxy NGC 7331 \citep{freedman01} as adopted by \citet{milisavljevic15}.

\section{Observations and Results}

\subsection{AMI Large Array} 

High cadence radio observations of SN 2014C were conducted by the AMI Large Array as part of a larger radio transient follow-up programme \citep[for example see][]{staley13,anderson14,fender15}. The AMI Large Array (which we will now refer to as AMI) is an eight antenna interferometer with baseline lengths between $18-110$~m, yielding a primary beam angular resolution of $5.5$ arcmin and $\approx40"$ arcsec, respectively, at 15.7 GHz. It measures a single polarisation ($I+Q$) and has an effective bandwidth of $13.9-17.5$ GHz corresponding to a flux root-mean square (rms) sensitivity of 3.3 mJy s$^{-1/2}$ \citep{zwart08}.

\begin{figure*}
\begin{center}
\includegraphics[width=\textwidth]{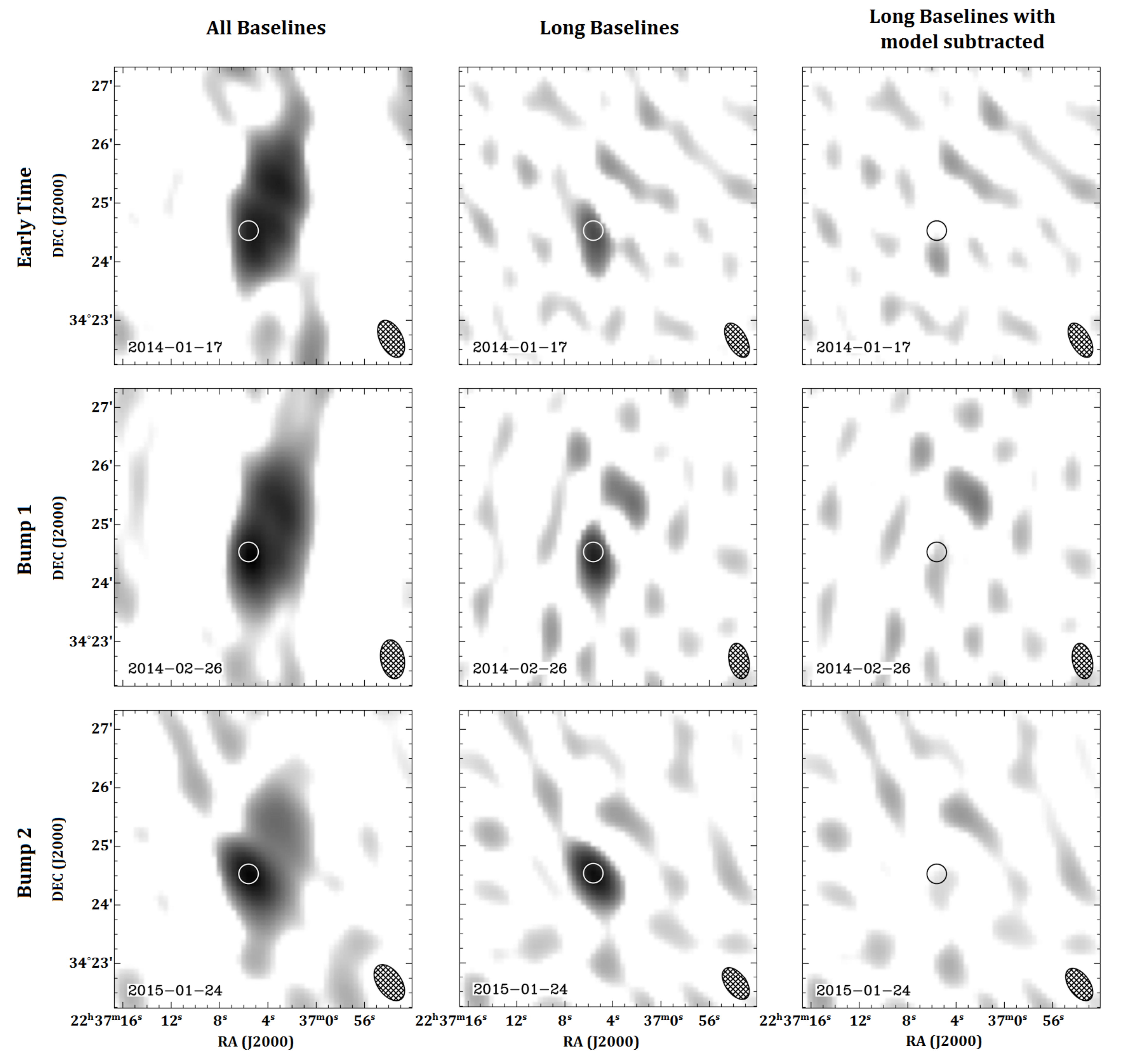}
\end{center}
\caption{AMI Large Array observation of SN 2014 taken at early times shortly after discovery on 2014 Jan 17 (56675 MJD - first row), on 2014 Feb 26 (56714 MJD - second row) during bump 1, and on 2015 Jan 24 (57047 MJD - third row) near peak brightness during bump 2 (see discussion of bumps in Section 3). The grey scale is logarithmic between each image's RMS and the brightest source flux of $\sim6$, $\sim8$ and $\sim30$ mJy for each of the three observations, respectively. The position of SN 2014C is indicated by the white circle (black circle for the third panel on each row) with a radius of $10''$. The beam is displayed in the bottom right hand corner of each image. The first panel for each observation shows the full baseline AMI image of SN 2014C, which is surrounded by resolved radio emission associated with the host galaxy NGC 7331. The second panel shows these AMI observations where only the $uv-$wavelengths $>2355\lambda$ have been imaged (i.e. long baselines) in order to minimise the resolved emission from the host and therefore contamination to SN 2014C. The third panel is the same as the second but with a point source model subtracted at the position of the SN to demonstrate that there is little contribution from the host galaxy when the shortest baselines have been removed from the analysis.}
\label{fig:2}
\end{figure*}

Observations of SN 2014C with AMI began on 2014 Jan 15, about 13 days after the earliest optical detection \citep{zheng14}. We then continued to observe the source nearly daily for $\sim1$ hour until 2014 Jan 19, and then every few days until 2014 Feb 2. The cadence then dropped to approximately once per week and continued until 2015 July 20, when AMI was shut down for a correlator upgrade. During this period we accumulated 82 AMI observations of SN 2014C totalling $\sim4$ days of observing time.

\tablefirsthead{\hline Obs. & Start & $\Delta t$ & Duration & Peak flux\\ date & (MJD)$^{\mathrm{~a}}$ & (days)$^{\mathrm{~b}}$ & (hrs) & density (mJy) \\ \hline}
\tablehead{\hline Obs. & Start & $\Delta t$ & Duration & Peak flux\\ date & (MJD)$^{\mathrm{~a}}$ & (days)$^{\mathrm{~b}}$ & (hrs) & density (mJy) \\ \hline}
\tabletail{\hline}
\tablelasttail{\hline \hline \multicolumn{5}{|l|}{$^{\mathrm{a}}$ Start date of AMI observation in MJD.} \\ \multicolumn{5}{|l|}{$^{\mathrm{b}}$ Days since predicted first light of SN 2014C;} \\ \multicolumn{5}{|l|}{2013 Dec 30 \citep[56656 MJD;][]{margutti16pp}.} \\ \hline}
\tablecaption{AMI 15.7 GHz flux density measurements for SN 2014C}
\label{tab:ami}
\begin{supertabular}{ccccc}
  2014-01-15  & 56672.68   & 16.68    &       1.0    &   $   2.78   \pm    0.29    $    \\       
  2014-01-16  & 56673.63   & 17.63    &       1.0    &   $   1.68   \pm    0.22    $    \\       
  2014-01-17  & 56674.58   & 18.58    &       1.0    &   $   1.38   \pm    0.25    $    \\       
  2014-01-18  & 56675.58   & 19.58    &       1.0    &   $   2.23   \pm    0.26    $    \\       
  2014-01-19  & 56676.53   & 20.53    &       1.0    &   $   1.52   \pm    0.33    $    \\       
  2014-01-20  & 56677.50   & 21.50    &       1.0    &   $   1.87   \pm    0.25    $    \\       
  2014-01-24  & 56681.57   & 25.57    &       1.0    &   $   2.11   \pm    0.31    $    \\       
  2014-02-01  & 56689.54   & 33.54    &       0.2    &   $   1.39   \pm    0.68    $    \\       
  2014-02-02  & 56690.50   & 34.50    &       1.0    &   $   1.86   \pm    0.32    $    \\       
  2014-02-07  & 56695.54   & 39.54    &       1.0    &   $   2.38   \pm    0.35    $    \\       
  2014-02-16  & 56704.58   & 48.58    &       1.0    &   $   3.59   \pm    0.28    $    \\       
  2014-02-22  & 56710.61   & 54.60    &       1.0    &   $   5.20   \pm    0.43    $    \\       
  2014-02-26  & 56714.54   & 58.54    &       1.0    &   $   5.27   \pm    0.35    $    \\       
  2014-03-05  & 56721.48   & 65.48    &       1.0    &   $   7.34   \pm    0.46    $    \\       
  2014-03-17  & 56733.42   & 77.42    &       1.0    &   $   6.39   \pm    0.47    $    \\       
  2014-03-23  & 56739.39   & 83.39    &       1.0    &   $   6.24   \pm    0.43    $    \\       
  2014-03-29  & 56745.39   & 89.39    &       1.0    &   $   6.66   \pm    0.46    $    \\       
  2014-04-05  & 56752.41   & 96.41    &       1.0    &   $   6.45   \pm    0.41    $    \\       
  2014-04-15  & 56762.38   & 106.37   &       1.0    &   $   5.50   \pm    0.50    $    \\       
  2014-04-27  & 56774.37   & 118.37   &       1.0    &   $   4.24   \pm    0.47    $    \\       
  2014-05-05  & 56782.45   & 126.45   &       1.0    &   $   5.90   \pm    0.41    $    \\       
  2014-05-13  & 56790.31   & 134.30   &       1.0    &   $   5.30   \pm    0.45    $    \\       
  2014-05-19  & 56796.28   & 140.28   &       1.0    &   $   5.40   \pm    0.40    $    \\       
  2014-05-30  & 56807.22   & 151.21   &       1.0    &   $   3.28   \pm    0.40    $    \\       
  2014-06-18  & 56826.20   & 170.20   &       1.0    &   $   5.62   \pm    0.56    $    \\       
  2014-07-04  & 56842.16   & 186.16   &       1.0    &   $   3.92   \pm    0.38    $    \\       
  2014-07-21  & 56859.19   & 203.19   &       1.0    &   $   4.56   \pm    0.58    $    \\       
  2014-07-27  & 56865.14   & 209.14   &       1.0    &   $   5.66   \pm    0.37    $    \\       
  2014-08-02  & 56871.11   & 215.11   &       1.0    &   $   6.87   \pm    0.45    $    \\       
  2014-08-07  & 56876.07   & 220.07   &       1.0    &   $   7.86   \pm    0.57    $    \\       
  2014-08-13  & 56882.03   & 226.03   &       1.0    &   $   9.78   \pm    0.59    $    \\       
  2014-08-18  & 56888.00   & 231.99   &       1.0    &   $   8.65   \pm    0.62    $    \\       
  2014-08-27  & 56896.01   & 240.01   &       1.0    &   $   9.88   \pm    0.62    $    \\       
  2014-09-04  & 56904.02   & 248.02   &       1.0    &   $   9.66   \pm    0.55    $    \\       
  2014-09-11  & 56911.00   & 255.00   &       1.0    &   $   11.42  \pm    0.70    $    \\       
  2014-09-13  & 56913.98   & 257.98   &       1.0    &   $   13.50  \pm    0.72    $    \\       
  2014-09-17  & 56917.95   & 261.95   &       1.0    &   $   12.81  \pm    0.81    $    \\       
  2014-09-24  & 56924.94   & 268.94   &       1.0    &   $   13.86  \pm    0.84    $    \\       
  2014-10-03  & 56933.88   & 277.88   &       1.0    &   $   13.26  \pm    0.81    $    \\       
  2014-10-10  & 56940.90   & 284.89   &       1.0    &   $   14.68  \pm    0.80    $    \\       
  2014-10-18  & 56948.99   & 292.99   &       1.0    &   $   15.33  \pm    0.82    $    \\       
  2014-10-26  & 56956.88   & 300.88   &       1.0    &   $   18.60  \pm    0.99    $    \\       
  2014-10-31  & 56961.81   & 305.81   &       1.0    &   $   19.18  \pm    1.06    $    \\       
  2014-11-06  & 56967.81   & 311.81   &       1.0    &   $   21.56  \pm    1.15    $    \\       
  2014-11-15  & 56976.78   & 320.78   &       1.0    &   $   21.71  \pm    1.15    $    \\       
  2014-11-25  & 56986.82   & 330.82   &       1.0    &   $   22.81  \pm    1.29    $    \\       
  2014-12-05  & 56996.77   & 340.76   &       1.0    &   $   22.45  \pm    1.21    $    \\       
  2014-12-11  & 57002.77   & 346.77   &       1.0    &   $   23.94  \pm    1.25    $    \\       
  2014-12-17  & 57008.73   & 352.73   &       1.0    &   $   22.30  \pm    1.16    $    \\       
  2014-12-19  & 57010.64   & 354.63   &       1.0    &   $   21.46  \pm    1.17    $    \\       
  2014-12-25  & 57016.71   & 360.71   &       1.0    &   $   25.16  \pm    1.37    $    \\       
  2015-01-02  & 57024.64   & 368.64   &       1.0    &   $   25.00  \pm    1.29    $    \\       
  2015-01-06  & 57028.61   & 372.61   &       1.0    &   $   24.61  \pm    1.28    $    \\       
  2015-01-14  & 57036.67   & 380.67   &       1.0    &   $   24.40  \pm    1.29    $    \\       
  2015-01-18  & 57040.65   & 384.65   &       1.0    &   $   25.60  \pm    1.36    $    \\       
  2015-01-24  & 57046.50   & 390.50   &       1.0    &   $   25.01  \pm    1.31    $    \\       
  2015-02-06  & 57059.50   & 403.50   &       1.0    &   $   27.17  \pm    1.45    $    \\       
  2015-02-13  & 57066.51   & 410.51   &       1.0    &   $   25.55  \pm    1.35    $    \\       
  2015-02-15  & 57068.52   & 412.52   &       1.0    &   $   27.73  \pm    1.44    $    \\       
  2015-02-21  & 57074.47   & 418.47   &       1.0    &   $   26.23  \pm    1.42    $    \\       
  2015-03-05  & 57086.52   & 430.52   &       0.6    &   $   24.58  \pm    1.28    $    \\       
  2015-03-05  & 57086.55   & 430.55   &       0.2    &   $   24.28  \pm    1.31    $    \\       
  2015-03-09  & 57090.34   & 434.34   &       1.0    &   $   25.12  \pm    1.30    $    \\       
  2015-03-19  & 57100.48   & 444.48   &       1.0    &   $   24.62  \pm    1.31    $    \\       
  2015-04-01  & 57113.45   & 457.45   &       1.0    &   $   19.89  \pm    1.04    $    \\       
  2015-04-11  & 57123.27   & 467.27   &       1.0    &   $   22.44  \pm    1.20    $    \\       
  2015-04-16  & 57128.37   & 472.37   &       1.0    &   $   21.94  \pm    1.15    $    \\       
  2015-04-23  & 57135.21   & 479.21   &       1.0    &   $   21.07  \pm    1.17    $    \\       
  2015-05-02  & 57144.30   & 488.30   &       1.0    &   $   22.12  \pm    1.20    $    \\       
  2015-05-04  & 57146.20   & 490.20   &       4.0    &   $   21.25  \pm    1.13    $    \\       
  2015-05-05  & 57147.22   & 491.22   &       4.0    &   $   21.54  \pm    1.24    $    \\       
  2015-05-11  & 57153.27   & 497.26   &       1.0    &   $   23.16  \pm    1.19    $    \\       
  2015-05-17  & 57159.28   & 503.28   &       1.0    &   $   21.59  \pm    1.15    $    \\       
  2015-05-23  & 57165.27   & 509.27   &       1.0    &   $   20.01  \pm    1.10    $    \\       
  2015-05-29  & 57171.27   & 515.27   &       1.0    &   $   20.14  \pm    1.12    $    \\       
  2015-06-05  & 57178.31   & 522.31   &       1.0    &   $   17.61  \pm    1.18    $    \\       
  2015-06-12  & 57185.18   & 529.18   &       1.0    &   $   19.66  \pm    1.12    $    \\       
  2015-06-25  & 57198.29   & 542.29   &       1.0    &   $   21.07  \pm    1.19    $    \\       
  2015-07-11  & 57214.25   & 558.26   &       1.0    &   $   19.64  \pm    1.03    $    \\       
  2015-07-18  & 57221.10   & 565.10   &       3.0    &   $   18.56  \pm    0.99    $    \\       
  2015-07-20  & 57223.09   & 567.09   &       3.0    &   $   18.51  \pm    1.02    $    \\       
\end{supertabular}

The reduction of the AMI observations was conducted using the fully-automated calibration and imaging pipeline \texttt{AMIsurvey} \citep{staley15}. For the initial reduction and calibration steps, \texttt{AMIsurvey} makes use of AMI dedicated software called \texttt{AMI-REDUCE} \citep{staley15b,davies09}. Phase calibration was conducted using short interleaved observations of J2248+3718 \citep[for further details on the absolute flux calibration of AMI Large Array see][]{franzen11}. For the imaging step, \texttt{AMIsurvey} then calls out to the automated  algorithm \texttt{chimenea}, built upon the Common Astronomy Software Applications package \citep[CASA;][]{jaeger08}, which is specifically designed for imaging multi-epoch radio transient observations \citep{staley15,staley15c}. During this step the shortest $uv-$distances $<45$m, corresponding to $<2355\lambda$ at 15.7 GHz, were not imaged in order to minimise any resolved radio emission from the host galaxy NGC 7331 (largest AMI angular size of $\approx3$ arcmin) contributing to the flux at the position of SN 2014C. This made us sensitive to only those angular scales between $\approx40$ arcsec and $\approx1.5$ arcmin at 15.7 GHz. Example images of three AMI observations of SN 2014C, taken at various levels of radio brightness, both before and after the shortest $uv-$distances had been subtracted can be found in Figure~\ref{fig:2}. Figure~\ref{fig:2} also includes a panel where a point source model at the position of SN 2014C has been subtracted to illustrate that there is little evidence of diffuse contamination. Therefore the removal of $uv-$wavelengths $<2355\lambda$ allows us to obtain a reasonable estimate of the flux from SN 2014C.

The radio flux densities of SN 2014C were measured from the short base-line subtracted images using the CASA task \texttt{imfit} and are reported in Table~\ref{tab:ami}. The flux errors correspond to the flux density error output by CASA \citep[based on Gaussian statistics as calculated by][]{condon98} added in quadrature to the 5 per cent systematic flux calibration error of AMI \citep{franzen11}. The SN 2014C AMI 15.7 GHz radio light curve is shown in Figure~\ref{fig:1} with detections spanning $\sim17$ to $\sim567$ days after the predicted first light. We were unable to measure the in-band radio spectral index with useful accuracy due to large, correlated uncertainties in the channel flux densities.

\subsection{eMERLIN}

High-resolution radio observations of SN 2014C were made using the electronic Multi-Element Remotely Linked Interferometer Network (eMERLIN). eMERLIN remotely connects to 5 telescopes that are operated in conjunction with telescopes at Jodrell Bank Observatory located in Cheshire, UK, with maximum baseline lengths of up to 217 km.\footnote{http://www.e-merlin.ac.uk/} For these observations the array was comprised of the Jodrell Bank Mark II, Knockin, Defford, Pickmere, and Darnhall 25 metre antennas and a 32-metre antenna at Cambridge (Jodrell Bank Mark II was not available for the eMERLIN observation in 2014).

The first eMERLIN observation of SN 2014C was conducted very early after discovery on 2014 January 19, 20 days following the predicted first light, using the C-band receiver tuned to a central frequency of 5.5 GHz. A further three observation during April and May 2015 were also obtained as part of director's discretionary time using L-band and C-band receivers. For these observation the instrument was tuned to central frequencies of 1.5 and 5.1 GHz respectively. All observations used a total bandwidth of 512 MHz. Data were processed and calibrated using the e-MERLIN pipeline \citep{argo15}, Parseltongue \citep{kettenis06}, and the Astronomical Image Processing System (AIPS) package produced by the National Radio Astronomy Observatory (NRAO). Data were initially edited with \texttt{SPFLG} and \texttt{IBLED} and averaged to 128 channels per intermediate frequency before further editing. \texttt{FRING} was initially used to derive the delay and rates of a bright calibrator and \texttt{CALIB} was used to solve for time-dependent phase and then amplitude variations. Flux calibration was performed using a short observation of 3C 286 and the flux density scale was set using \texttt{SETJY} \citep{perleybutler13}. Bandpass calibration was then performed using the bright source 1407+284, and the phase calibration was conducted using J2248+3718 ($\alpha \mathrm{(J2000.0)} =  22^{\mathrm{h}}48^{\mathrm{m}}37\overset{\mathrm{s}}{.}9110$ and $\delta (\mathrm{J2000.0}) = +37^{\circ}18'12\overset{''}{.}468$), which has an integrated flux density of 498 mJy at 5.5 GHz.

\begin{figure}
\begin{center}
\includegraphics[width=0.5\textwidth]{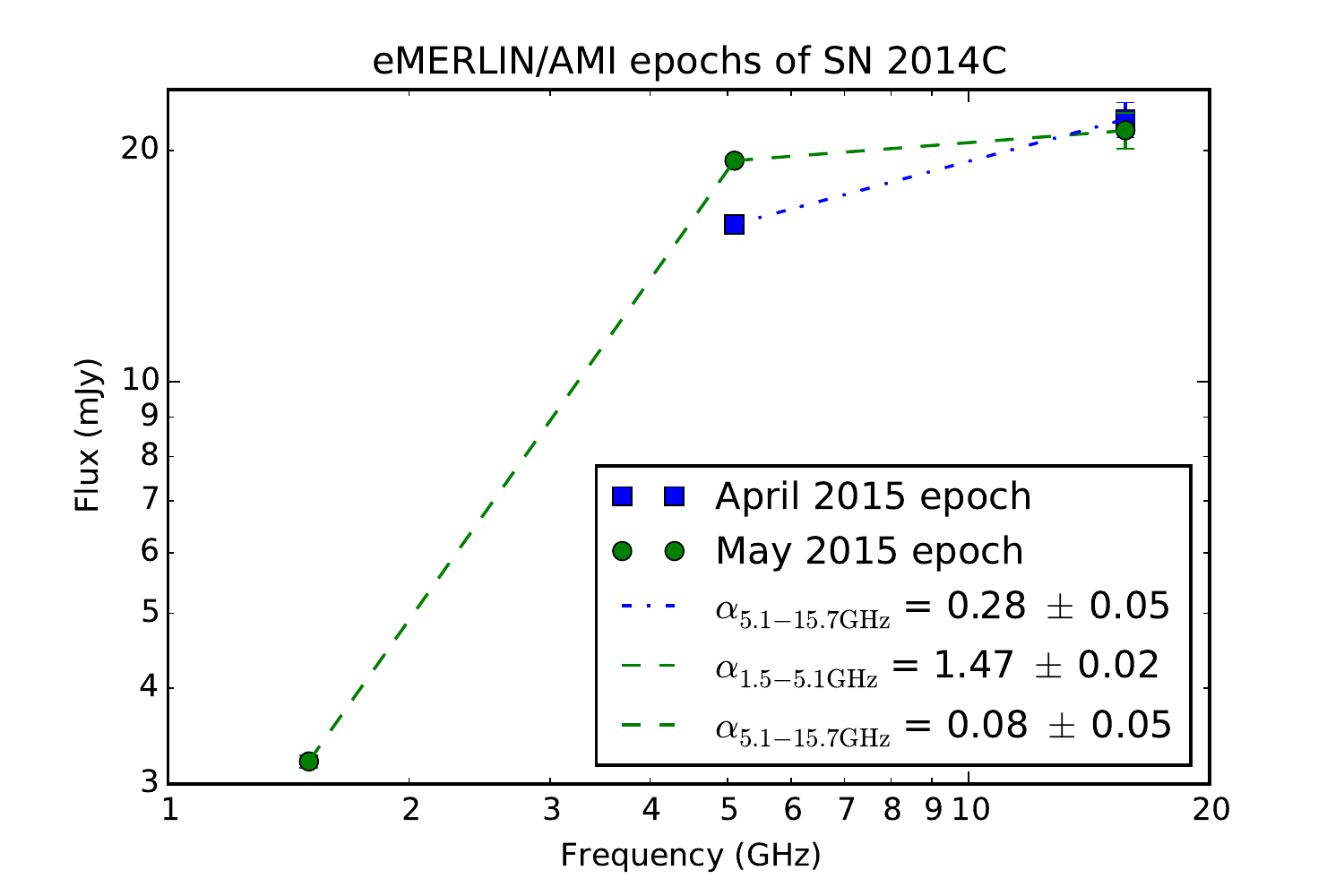}
\end{center}
\caption{The radio spectrum of SN 2014C during the two 2015 eMERLIN epochs. The eMERLIN $1\sigma$ error bars are encompassed within the plotted data points. Note that the AMI 15.7 GHz flux for both epochs are very similar resulting in the data points overlapping. The lines connecting the adjacent spectral bands are to aid the eye and are not necessarily representative of the true spectral indices.}
\label{fig:3}
\end{figure}

The initial eMERLIN observation in January 2014 did not detect SN 2014 with a $3\sigma$ upper-limit of $243\mu$Jy but was detected in the 2015 observations. The fluxes measured by eMERLIN in 2015 (averaged across each band in Table~\ref{tab:emerlin}) were divided into two epochs, April and May 2015 (as the 1.5 and 5.1 GHz observations in May were taken within 2 days of each other), and compared to the closest AMI observations taken in time (within $\sim1$ day of the eMERLIN observations). The resulting spectra for these two epochs can be found in Figure~\ref{fig:3}. The spectral index between each adjacent observing bands are also indicated in Figure~\ref{fig:3} (note that these lines just connect the points and may not be representative of the true spectral indices). The 5.1 and 15.7 GHz fluxes are similar, with both being comparable in the second epoch indicating that the spectrum transitioned from optically thick to optically thin for the second time between these two frequencies at $\sim490$ days post-explosion. 

The 2015 observations showed SN 2014C to be unresolved at the eMERLIN resolution of 40 mas. The position errors, which are due to the phase transfer from the reference source, are about 4 mas at 5.1 GHz. Relative to the phase calibrator position we found SN 2014C to be located at $\alpha \mathrm{(J2000.0)} =  22^{\mathrm{h}}37^{\mathrm{m}}05\overset{\mathrm{s}}{.}6042$ ($\pm 0.27$s) and $\delta (\mathrm{J2000.0}) = +34^{\circ}24'31\overset{''}{.}424$ ($\pm 4$mas).

\begin{table}
\begin{center}
\caption{eMERLIN observations of SN 2014C}
\label{tab:emerlin}
\begin{tabular}{lccccc}
\\
\hline
Obs. & Start & $\Delta t$ & Duration & $\nu$ & Flux\\
Date & (MJD) & (days) & (hrs) & (GHz) & (mJy) \\
\hline
2014-01-19 & 56676 & 20 & 4.0 & 5.5 & $<0.243$ \\
2015-04-17 & 57129 & 473 & 18.0 & 5.1 & $16.03 \pm 0.05$ \\
2015-05-03 & 57145 & 489 & 15.75 & 1.5 & $3.21 \pm 0.06$ \\
2015-05-05 & 57147 & 491 & 18.0 & 5.1 & $19.40 \pm 0.07$ \\
\hline
\end{tabular}
\end{center}
The errors are $1\sigma$ and the observation on 2014-01-19 gives a $3\sigma$ upper-limit. \\
\end{table}

\subsection{Keck Optical Spectra}

\begin{figure*}
\begin{center}
\includegraphics[width=1.0\textwidth]{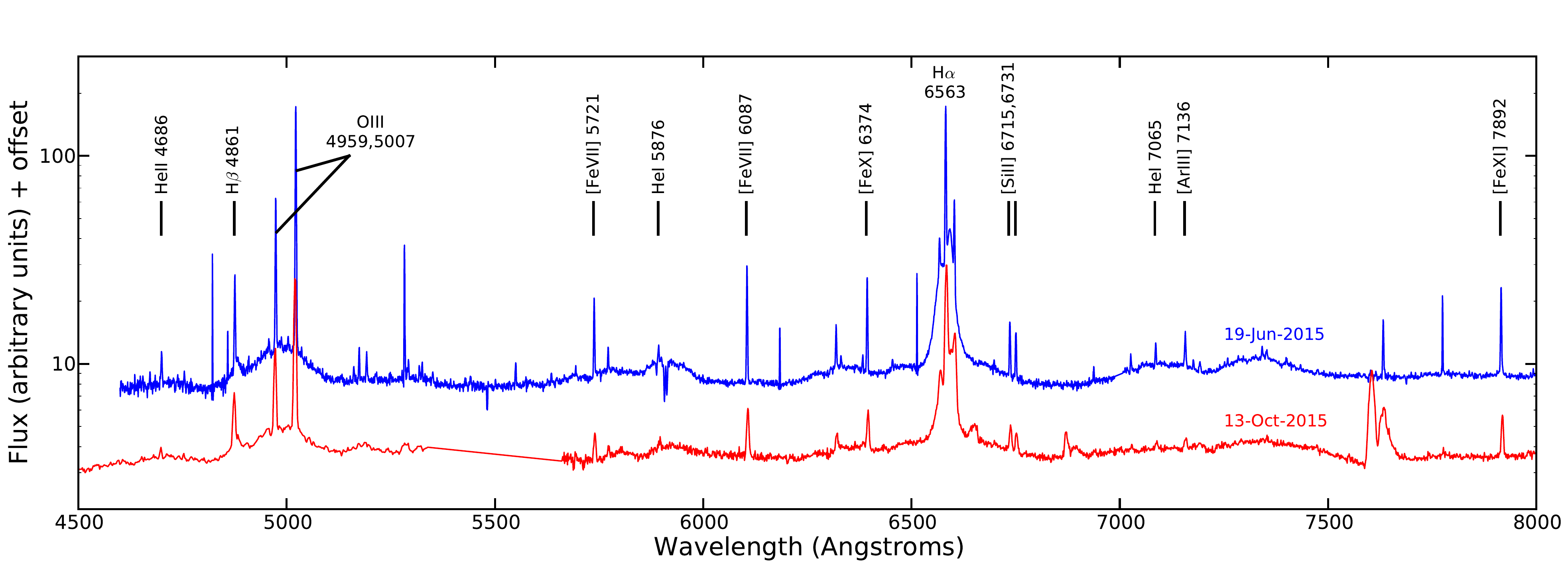}
\end{center}
\caption{The two Keck optical spectra taken on 2015 June 19 (blue/upper) and 2015 Oct 13 (red/lower), 536 and 652 days post-explosion, respectively. The 2015 Oct 13 spectrum has not been corrected for telluric absorption.}\label{fig:4}
\end{figure*}

In order to investigate the shock velocities and properties of the ambient medium, we carried out optical follow up observations of SN 2014C. We observed with the DEIMOS spectrograph \citep{faber03} on the Keck II telescope on 2015 June 19 (57192 MJD; 536 days post-explosion), after SN 2014C was visible again in the night sky. Another spectrum was obtained on 2015 Oct 13 (57308 MJD; 652 days post-explosion) with the LRIS spectrograph \citep{oke95,rockosi10} on the Keck I telescope. Both spectra are shown in Figure~\ref{fig:4}. These spectral epochs closely follow the time of the peak of the second bump seen in our radio light curve as indicated in Figure~\ref{fig:1} and the emission line profiles match \citet{milisavljevic15}'s late-time (373 days post-explosion) spectral classification of SN 2014C as a Type IIn (see their Figure 3). 

\begin{figure*}
\begin{center}
\includegraphics[width=0.45\textwidth]{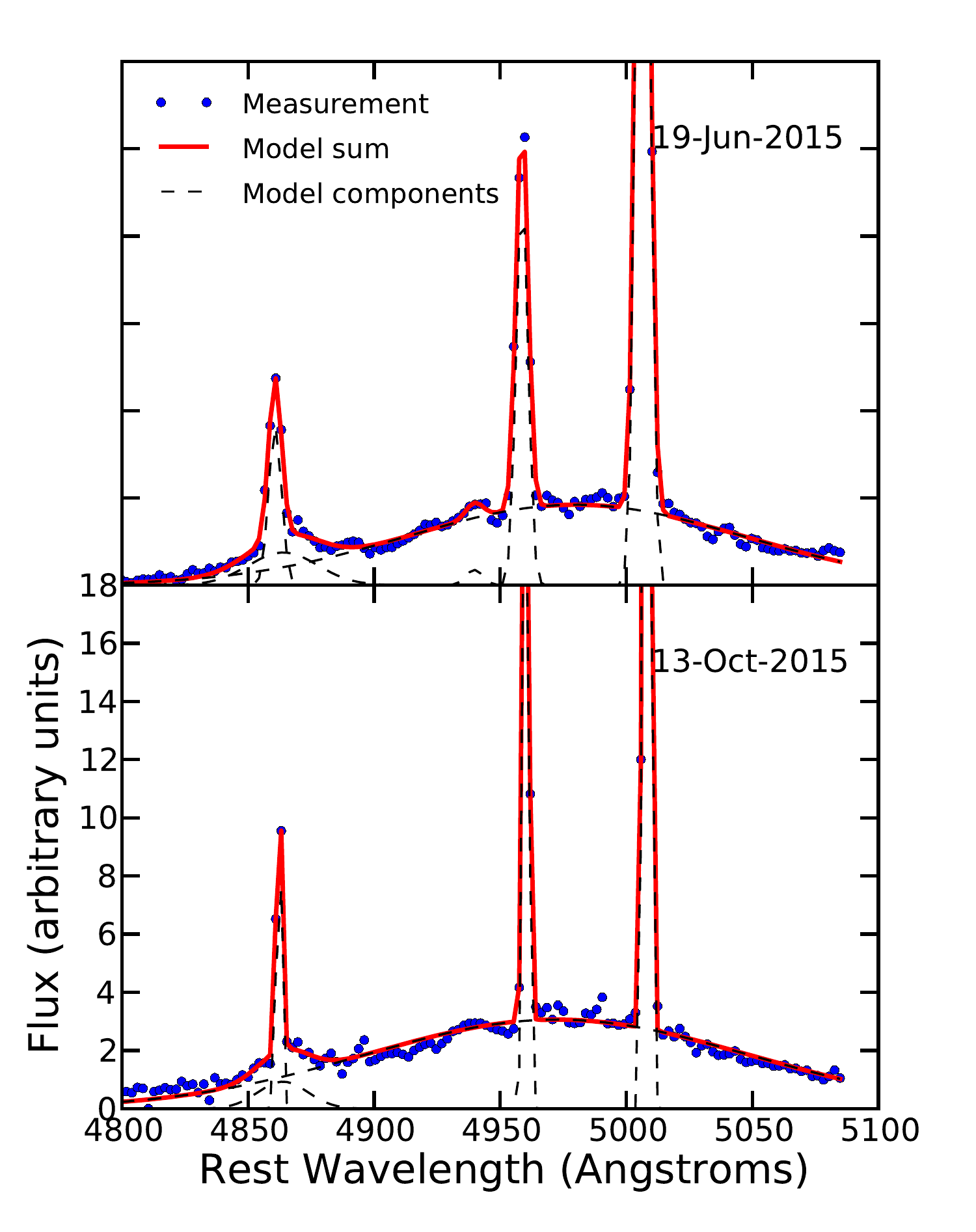}
\includegraphics[width=0.45\textwidth]{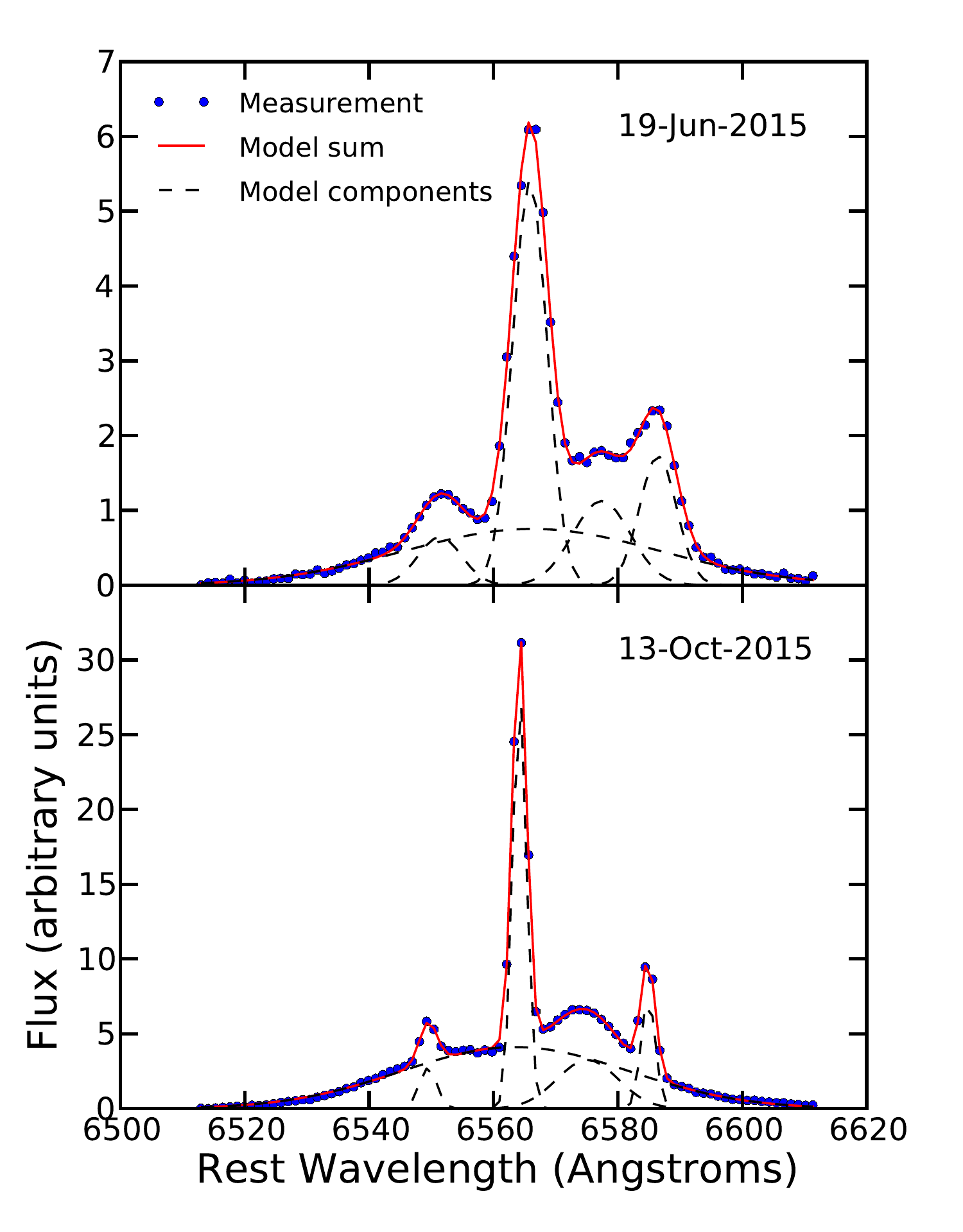}
\end{center}
\caption{Gaussian emission line fitting to the complexes surrounding [O{\sc iii}] $\lambda\lambda4959,5007$ (left) and H$\alpha$ (right) in the 2015 June 19 and 2015 Oct 13 epoch Keck spectra.}\label{fig:5}
\end{figure*}

The Keck spectra reveal narrow emission lines superimposed with broad components, especially [O{\sc iii}] $\lambda\lambda4959,5007$ and H$\alpha$. 
These broad components were then fit with Gaussian components using Python/Scipy least-square fitting algorithms. This resulted in (FWHM) velocities of 940 km s$^{-1}$ and 840 km s$^{-1}$ ($\pm 100$ km s$^{-1}$) for H$\alpha$, and 4020 km s$^{-1}$ and 4580 km s$^{-1}$ ($\pm 500$ km s$^{-1}$) for [O{\sc iii}] for the 2015 June 19 and 2015 Oct 13 spectra, respectively (see line fitting in Figure~\ref{fig:5}). Our Gaussian fit to the [O{\sc iii}] line agrees with the \citet{milisavljevic15} lower-limit of $\gtrsim3500$ km s$^{-1}$ at 481 days post-explosion (467 days post-maximum light).

SN [O{\sc iii}] and H$\alpha$ broad emission lines are the optical signatures originating from the stellar ejecta becoming heated and ionised as the reverse shock (RS) propagates upstream \citep{chevalier94}. As mentioned by \citet{milisavljevic15}, broad [O{\sc iii}] usually only begins to appear many years following the explosion \citep{milisavljevic12}, when the RS interacts with the oxygen-rich stellar ejecta. Our early detections of [O{\sc iii}] indicates that this strong interaction between the SN and the CSM has occurred much earlier for SN 2014C as the ejecta reaches the proposed H-rich shell \citep{milisavljevic15,margutti16pp}. The broad H$\alpha$ emission component is typical of SN IIn as it originates from shocked or photoionized SN material resulting from the ejecta colliding with the dense H-rich material \citep{chugai94}. Our late-time spectra demonstrate the continued broadening in H$\alpha$ emission up to $\gtrsim4000$ km s$^{-1}$, from 1200 km s$^{-1}$ obtained 113 days post-maximum light \citep{milisavljevic15}, indicating the continued interaction of the SN with the H-rich shell.

\section{The radio light curve evolution of SN~2014C}

The AMI light curve for SN 2014C shows unusual morphological behaviour in the form of two distinct peaks (see Figure~\ref{fig:1}). Late-time radio rebrightening and deviations from a typical light curve evolution have been seen in several SNe but the distinct double peaked nature is quite unusual. The most direct analogues of the radio behaviour we observe in the AMI light curve of SN 2014C are the Type Ic SN 2007bg \citep{salas13} and the Type Ic broad-line SN PTF11qcj \citep{corsi14}. We adopt the same analysis strategy as \citet{salas13} where we divide the light curve into distinct stages. We first discuss the early time emission and then divide the light curve into SN bump 1 and SN bump 2 to explore the unusual double peaked behaviour. 

\subsection{Early time emission} 

This stage encompasses the initial early time (within $\approx 34$ days following first light) radio detections of SN 2014C. These first eight detections, represented by red stars in Figure~\ref{fig:1}, precede the power-law rise that we call bump 1 (depicted in blue). While these first 8 detections appear to vary, a weighted mean fit to the data shows it to be insignificant, with a probability of $92\%$ that any deviation is due to random chance. However, the 7.1 GHz light curve of SN 2014C presented by \citet{margutti16pp} in their Figure~9 also shows the flux to vary at early times ($<20$ days) so the contemporary behaviour AMI observes at 15.7 GHz may have some real element. SN 2007bg shows variability at early times and \citet{salas13} suggests this could be due to interstellar scintillation or a real deviation from pure synchrotron self-absorption (SSA) due to clumpy CSM. 

In order to investigate whether interstellar scintillation is a factor at these early times we used the NE2001 model of the distribution of free electrons in our Galaxy \citep{cordes02} to determine the transitional frequency at which the scattering strength is unity ($\nu_{0}$=10 GHz) and the corresponding angular size limit \citep[$\theta_{\mathrm{FO}}=2.1 \times 10^{4} SM^{0.6} \nu_{0}^{-2.2} \approx 1\mu$as, where $SM$ is the scattering measure;][]{granot14} for an extragalactic source at the position of SN 2014C. Our observations at 15.7 GHz are therefore in the weak scattering regime. If we assume a uniform minimum shock expansion velocity of 13000 km s$^{-1}$ \citep[the bulk ejecta velocity measured by][]{milisavljevic15},\footnote{Note that the leading ejecta that are responsible for the radio emission are likely to be 2 or 3 times faster than the bulk flow.} and that the source expands isotropically, then by the time of our first observation at 17 days post-explosion we estimate that the source size is $\theta_{\mathrm{S}}\approx17 \mu$as. Following \citet{walker98} formalism and Table 1 of \citet{granot14}, the modulation index is $m_{\mathrm{w}}=0.01$, with scintillation acting on a time scale of 34 hrs. Given the small modulation index it is highly unlikely that scintillation is causing any flux variability at the AMI observing frequency after 17 days. The deviation of the SN 2014C light curve from typical radio SN turn-on behaviour is therefore likely due to a clumpy CSM or contaminated by resolved emission from the host that may not have been properly removed during the reduction step.

We also obtained a $3\sigma$ upper-limit of 0.243 mJy at 5.5 GHz with eMERLIN just 20 days after the predicted first light. A 5 GHz detection was reported by \citet{kamble14} at 12 days post-explosion, but no flux measurement was given.  Reprocessing the data from this observation (project code: 13A-370) we find a weak compact component with a peak of 0.18 mJy/beam, putting it below the detection threshold in the e-MERLIN data at the time of observation.

\subsection{SN Bump 1}

At $\sim34$ days, when SN 2014C is classified as a SN Ib \citep{milisavljevic15}, the radio light curve begins to follow a typical SN radio turn-on. The brightest flux measured at 15.7 GHz for bump 1 occurred at 65.5 days (56721.5 MJD) resulting in a peak spectral luminosity of $1.9 \times 10^{27}$ erg s$^{-1}$Hz$^{-1}$.\footnote{All spectral luminosities are bandpass corrected given the SN redshift but are not $k-$corrected.} This peak, while consistent in luminosity, occurs at a much later time than most SNe Ibc when compared to Figure~2 of \citet[][plot of the peak radio luminosity vs the time of peak normalised to 5 GHz]{chevalier06a}. It is more consistent with the peak time and generally fainter fluxes of SNe IIb and IIL and indicates shock velocities between $1000-10000$ km s$^{-1}$ if we assume that SSA is the dominant absorption mechanism. This implied velocity is smaller than the bulk velocity reported by \citep{milisavljevic15} but the radio emitting region must be going at least as fast as the expanding photosphere. This means that SSA may not be the only, or even dominant, absorption mechanism and that there must be a denser or clumpier surrounding CSM than what is usually seen for SNe Ibc. 

An optical spectrum of SN 2014C was obtained 127 days after the predicted first light, towards the end of bump 1, when SN 2014C became observable at night again \citep{milisavljevic15}. This spectrum showed that SN 2014C had metamorphosed into a SN IIn as it showed a bright H$\alpha$ line that was likely dominated by emission from the SN-CSM interaction. \citet{milisavljevic15} concluded that ``an extraordinary event must have occurred while SN 2014C was hidden behind the Sun'', between February and May 2014 (days $\approx 35 - 145$ post-first light), which covers the entire evolution of bump 1 seen in the AMI light curve. The AMI observations therefore confirm this conclusion as around day $\sim100$ the flux becomes unsteady, deviating from the expected decline usually seen in the optically thin regime (see Section 4.1) and varying by a factor of up to $\sim2$ as it approaches day $186$. Such variations are likely due to density enhancements in the CSM as the shock approaches the higher density medium (see Section 3.3), which was a feature also seen in the 15 and 22 GHz light curves of SN 2007bg \citep{salas13}. However, at day $186$ the overall flux decline stopped, stabilised and began to rebrighten (the start of bump 2 - see Section 3.3). 

\subsection{SN bump 2}

Following this period of discontinuity the flux begins to steadily increase again for a late-time rebrightening, once again following the typical power-law rise associated with optically thick emission (see Section 4.1). The radio rebrightening is also temporary coincident with the observed IR brightening reported by \citep{tinyanont16pp}. The brightest measured flux of bump 2 at $\sim412.5$ days (57068.5 MJD) was a factor of $\sim4$ times more luminous than the brightest flux measurement near the peak of bump 1, which is a much larger difference than was observed for both SN 2007bg and SN PTF11qcj \citep{salas13,corsi14}. This corresponds to a peak spectral luminosity of $7.2 \times 10^{27}$ erg s$^{-1}$ Hz$^{-1}$, which is consistent with the peak luminosities seen from most known SNe \citep[see Figure 2 of][]{perez-torres15}. However, the late time of this peak is particularly consistent with other known SNe IIn, and therefore reflects the optical spectral classifications obtained by the Keck spectra and \citet{milisavljevic15} several hundreds of days post-explosion. While SN 2014C is in the intermediate luminosity range when compared to other SNe IIn, it is one of the latest peaking radio SNe of any type, including SN 2007bg \citep[see Figure 2 of][]{perez-torres15}, particularly when considering the high frequency observing band of 15.7 GHz. It is also brighter than all of the radio SNe Ibc with the exception of SN 2007bg and the relativistic SNe \citep[some of which are associated with GRBs, see][]{soderberg10nat}. 

Closely following the peak time of bump 2 we obtained two epochs of eMERLIN observations at 1.5 and 5.1 GHz (see Table~\ref{tab:emerlin}). As mentioned in Section 2.2, the second eMERLIN epoch around 490 days post-explosion showed that SN 2014C was almost about to turn optically thin around 5.1 GHz (the eMERLIN flux measurements have also been included in the AMI light curve in Figure~\ref{fig:1}). The spectral index between the eMERLIN 1.5 and 5.1 GHz bands during this epoch is not as steep as is usually expected from SSA or FFA but this could be due to lower frequency contributions to the flux from bump 1. The broad spectral lines in our Keck spectra also indicate that at late times ($>500$ days post-explosion) the SN is strongly interacting with a dense CSM.

\begin{figure}
\begin{center}
\includegraphics[width=0.45\textwidth]{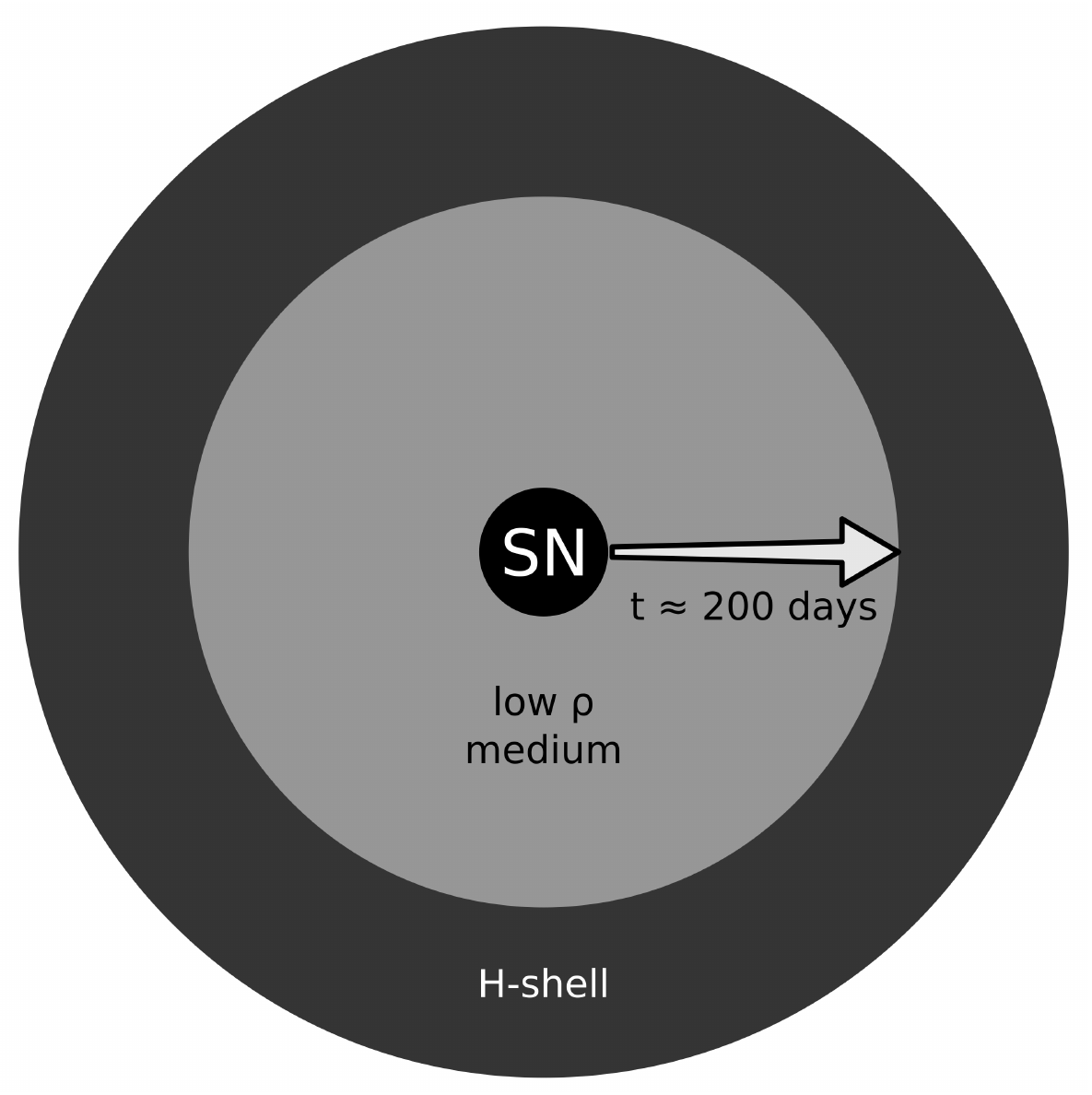}
\end{center}
\caption{Schematic of the CSM density we prepose is surrounding SN 2014C (inner filled back circle) based on the AMI light curve morphology. The darker regions indicate areas of higher CSM density. The white arrow represents the isotropically expanding SN shock wave as it first propagates through the low density ($\rho$) medium producing bump 1, and then impacts the denser H-shell at $\sim200$ days post-explosion, the start of bump 2. This schematic is not to scale and is provided purely to aid discussion in the text.}\label{fig:6}
\end{figure}

Using the AMI light curve morphology we suggest the following structure of the environment surrounding SN 2014C, which is illustrated in the schematic shown in Figure~\ref{fig:6} where darker regions indicate areas of higher CSM density. At very early times the SN shock wave is possibly interacting with a clumpy CSM. After $\sim34$ days post-explosion the light curve begins to increase steadily as is expected from a shock wave interacting with a low density and possibly steadily decreasing CSM (with a density falling off as $1/R^{2}$), peaking around $\sim80$ days post-explosion. However, around $\sim100$ days post-burst the light curve begins to vary and deviate from the power-law decline as the shock wave likely encounters a more inhomogeneous CSM near the edges of the H-shell proposed by \citet{milisavljevic15} and \citet{margutti16pp}. By $\sim200$ days the SN shock wave has hit the H-shell as the radio emission begins to steadily rise for the second time as the electrons are accelerated again due to the sudden density increase.

\section{Light curve analysis of SN 2014C}

\subsection{AMI light curve modelling}

In this section we model the AMI 15.7 GHz light curve of SN 2014C. The radio emission from SNe arises from nonthermal synchrotron emission generated by a population of relativistic particles such that the energy spectrum is $N(E) = N_{0}E^{-p}$, where N is the density of the relativistic particles and $p$ is the electron spectral index \citep{chevalier98}. These particles are accelerated in the shock region between the SN blast wave and the dense CSM that was ionised by the initial SN optical/UV/X-ray flash \citep{vandyk94}. This CSM was ejected by the progenitor star via its stellar wind or through more extreme mass-loss episodes during its lifetime. The particles then gyrate in the post-shock magnetic field producing synchrotron emission that is detectable at radio wavelengths\citep{chevalier82,vandyk94,chevalier98,weiler02,weiler07}. The radio emission is initially absorbed, usually due to either synchrotron self-absorption (SSA) from an internal medium or by free-free absorption (FFA) from the external CSM, and then rapidly rises in flux due to a decrease in the absorption processes as the radio-emitting region expands leaving less absorbing material that is either internal or along the line-of-sight \citep{vandyk94,weiler02,weiler07}. While the flux increases the emission is considered optically thick. It then peaks as it transitions from optically thick to optically thin and slowly declines in flux as the blast wave from the initial SN explosion decelerates as it interacts with the CSM \citep{chevalier98}.

A large amount of radio SN light curve modelling over the last 20 years has demonstrated that the early time radio emission from most SNe, particularly SNe Ibc, is primarily subject to internal SSA \citep{chevalier98,chevalier06}. However, the radio emission from SNe with extensive CSM (such as SNe IIn) can be well described by external FFA models, particularly at late times when the ejecta shock front has slowed and the ionised medium completely encompasses the emitting source \citep{weiler86,weiler90,weiler02}. \citep[Note that there could be other sources of absorbing media that could result in internal and other forms of external FFA absorption as described by][but given our limited dataset we will not explore these scenarios.]{weiler02} We will therefore model the AMI radio light curve of SN 2014C with a simple SSA and FFA model. In these models it is assumed that the CSM was created by a constant mass-loss rate ($\dot{M}$) due to a steady wind velocity ($v_{w}$) of the progenitor with the density dropping off smoothly as the inverse square of the radius R such that $\rho_{\mathrm{CSM}} \propto \dot{M}/(v_{w}R^{2})$ \citep{weiler07}. This is clearly not the case for SN 2014C, with its early time deviation from typical SN radio turn-on and the double bump morphology of its light curve. As the two bumps in the light curve are likely due to two different density regimes (and therefore mass-loss episodes) we will assume that the above condition holds for each bump and therefore model them separately.

The flux density evolution of radio SNe that are subject to SSA can be described by the parameterised model proposed by \citet{chevalier98} and adapted by \citet{weiler02,weiler07}:


\begin{align}\label{eq:1}
\begin{split}
S = K_{1}~\bigg(\frac{\nu}{5~\mathrm{GHz}}\bigg)^{\alpha} \bigg(\frac{t-t_{0}}{1~\mathrm{d}}\bigg)^{\beta} \bigg(\frac{1-\mathrm{e}^{-\tau_{SSA}}}{\tau_{SSA}}\bigg),
\end{split}
\end{align}
with the absorption expression being dominated by internal SSA where the optical depth ($\tau_{SSA}$) is given by:
\begin{align}\label{eq:2}
\begin{split}
\tau_{SSA} = K_{5}~\bigg(\frac{\nu}{5~\mathrm{GHz}}\bigg)^{\alpha-2.5} \bigg(\frac{t-t_{0}}{1~\mathrm{d}}\bigg)^{\delta''}.
\end{split}
\end{align}
In these equations $t_{0}$ is the explosion date, $K_{1}$ is the flux density normalisation, $K_{5}$ is the internal, non-thermal, SSA normalisation and $\delta''$ determines the time dependence of the optical depth for the SSA internal absorption. When modelling we solved for the latter three parameters.

\begin{figure*}
\begin{center}
\includegraphics[width=0.45\textwidth]{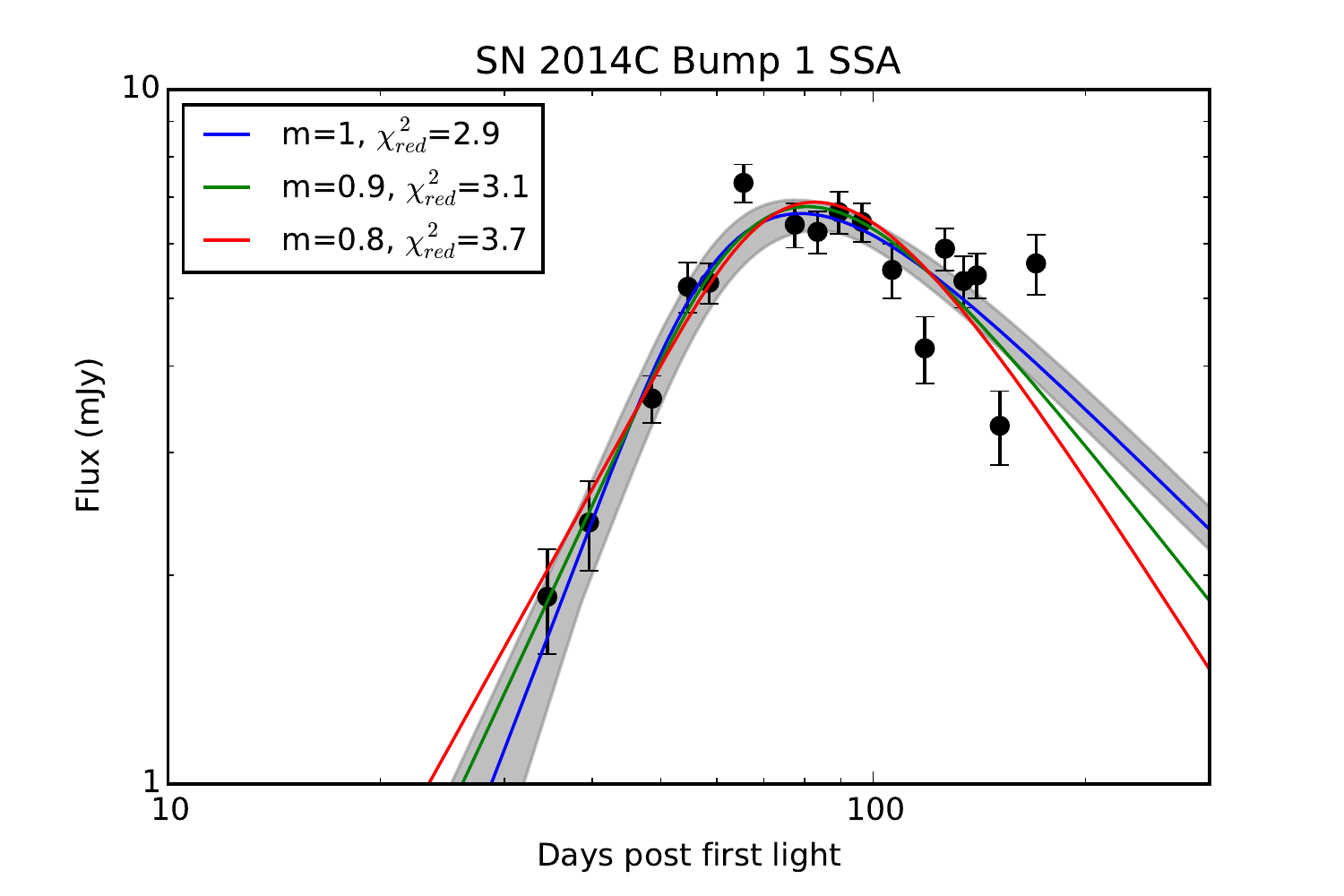}
\includegraphics[width=0.45\textwidth]{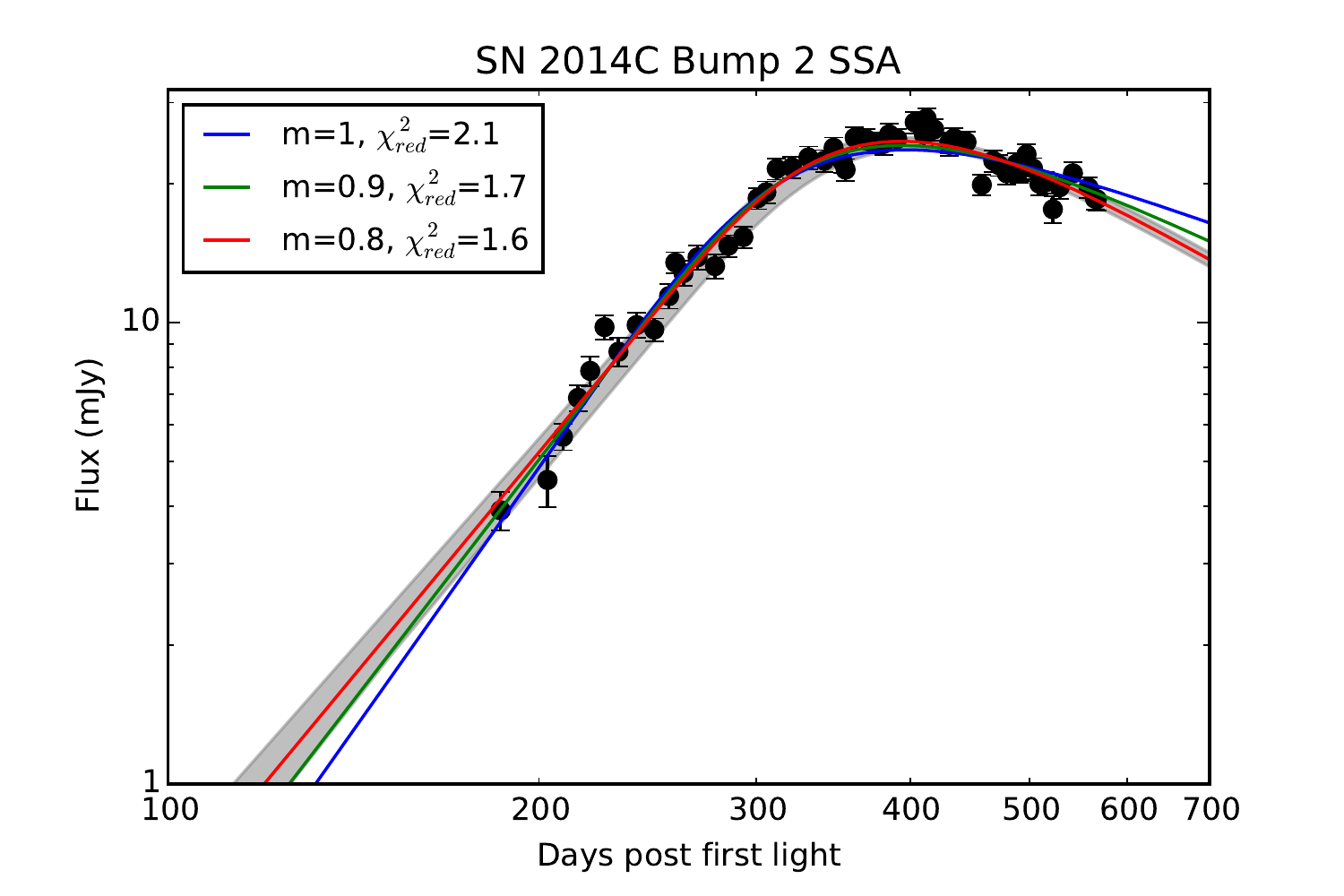}
\end{center}
\caption{Best fit SSA models to bump 1 (left panel) and bump 2 (right panel) for different deceleration parameters ($m$). The $\chi^{2}_{red}$ is provided for each of the three fits. The shaded grey area represents the $1\sigma$ error on the model fit corresponding to the best (lowest) $\chi^{2}_{red}$, which is $m=1$ and $m=0.8$ for bumps 1 and 2, respectively.}\label{fig:7}
\end{figure*}

\begin{figure*}
\begin{center}
\includegraphics[width=0.45\textwidth]{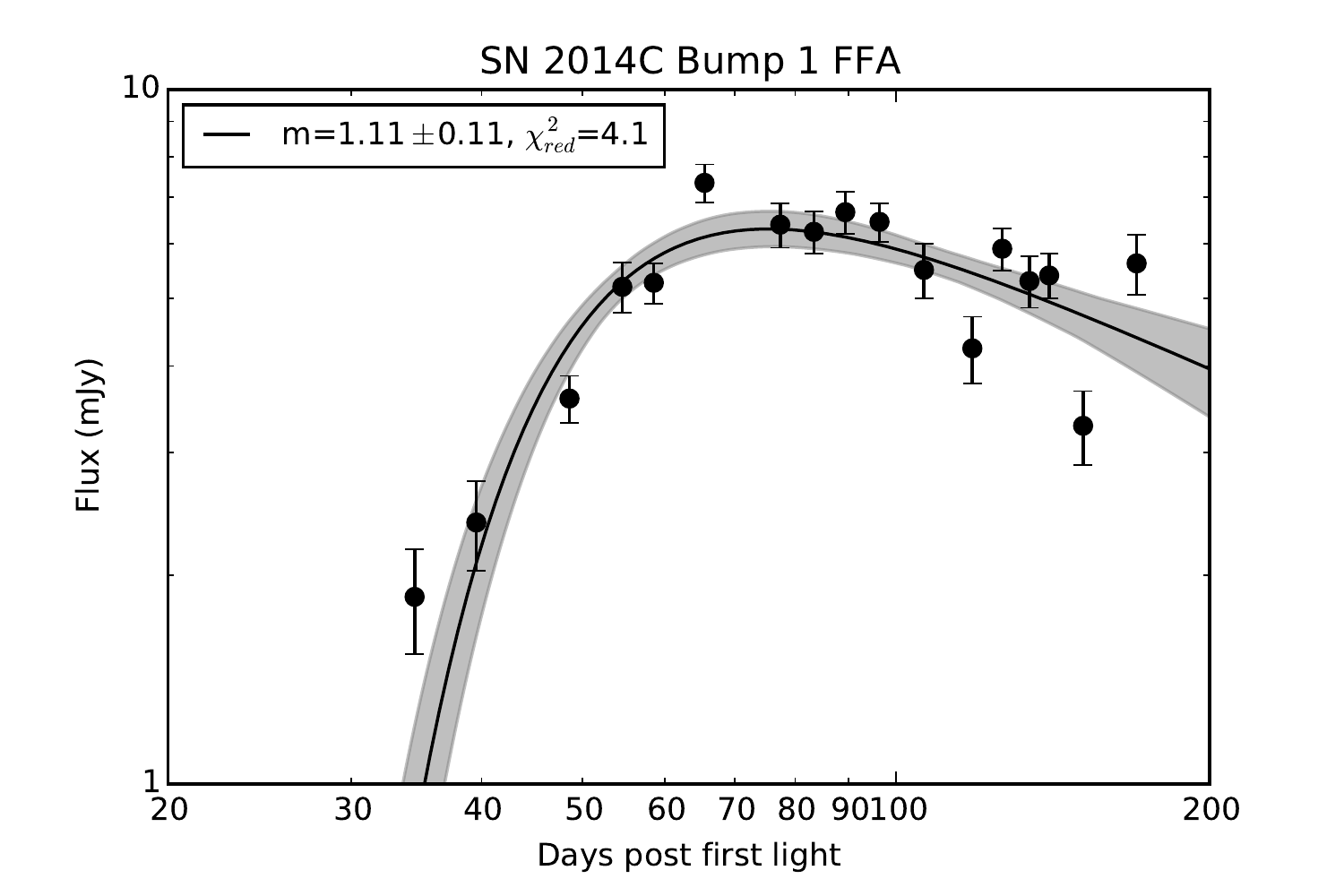}
\includegraphics[width=0.45\textwidth]{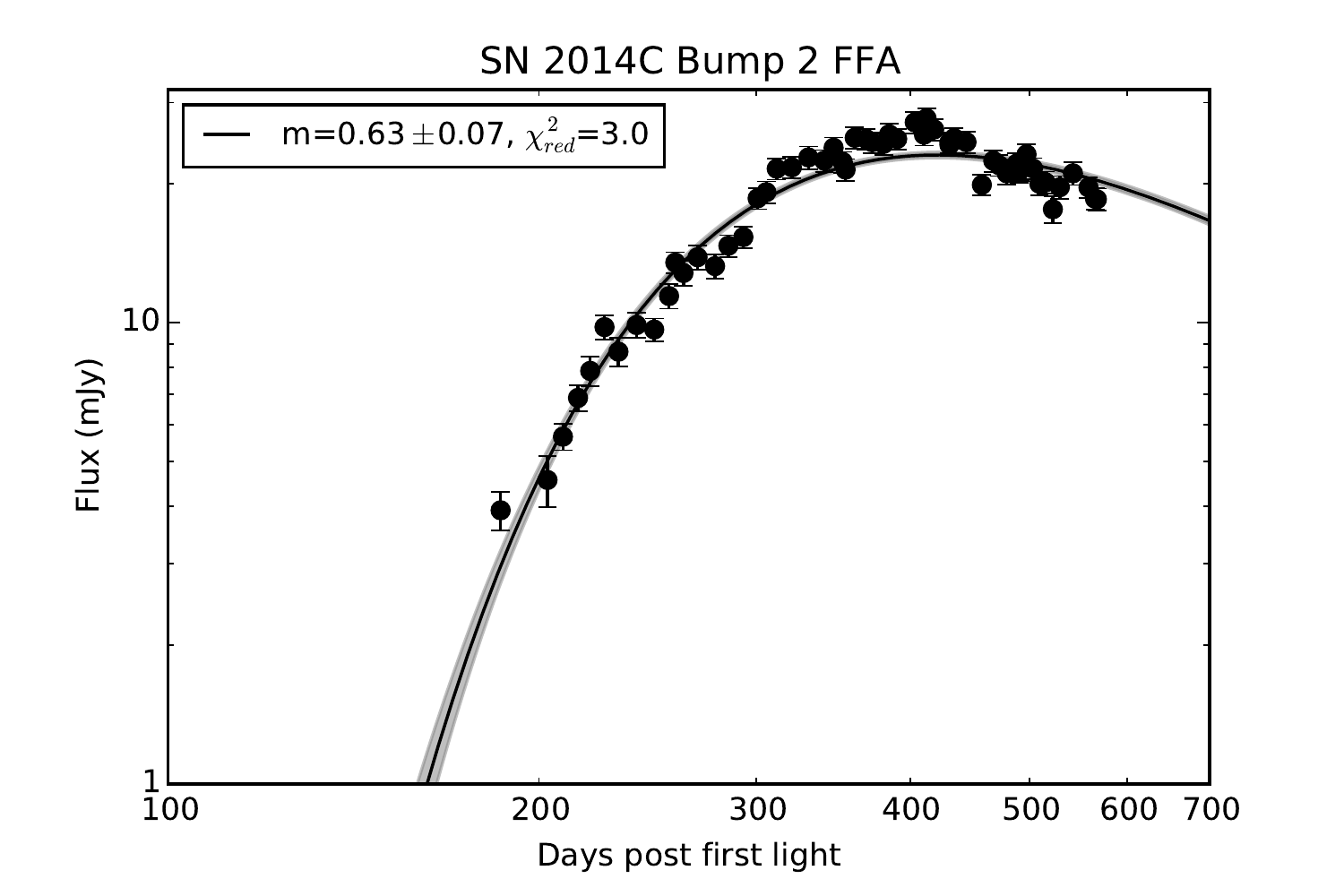}
\end{center}
\caption{Best fit FFA models to bump 1 (left panel) and bump 2 (right panel) assuming a spectral index of $\alpha=-0.7$. The fitted deceleration parameter $m$ and the resulting $\chi^{2}_{red}$ is provided for each fit. The shaded grey area represents the $1\sigma$ error on the model fit.}\label{fig:8}
\end{figure*}

In order to minimise the number of free parameters during the fitting process we made several assumptions about the spectral and temporal shape of the radio emission. The spectral index of a radio SN ($\alpha$ where $F_{\nu} \propto \nu^{\alpha}$) relates to the spectral index of the injected electron energy spectrum such that $p=-2 \alpha + 1$ \citep{weiler02}. Given that we are limited to a single frequency, we have fitted both bump 1 and bump 2 with the SSA model adopting the spectral index $\alpha \approx -1$, which is appropriate for SN Ibc \citep{weiler02,chevalier06}. This spectral index assumes an injected electron energy spectrum with $p=3$, which has been shown to be a reasonable assumption for SN Ibc \citep{chevalier06}. However, it should be noted that $p$ has also been shown to slowly vary over time \citep[e.g. SN 2011dh;][]{horesh13} and in the case of SN 2002ap have an initial electron energy injection as low as $p \approx 2$ \citep{bjornsson04,chevalier06}.

According to the \citet{chevalier82} model the blast wave radius increases as a power of time ($R \propto t^{m}$), where $m \leq 1$ is the deceleration parameter. The deceleration parameter is related to the density distribution index $n$ of the SN ejecta ($\rho_{ej} \propto r^{-n}$), such that $m=(n-3)/(n-2)$. By changing the value of the temporal index $\beta$, where $F_{\nu} \propto t^{\beta}$, we can probe for different deceleration parameters since $m = -(\alpha - \beta - 3)/3$ \citep[for cases where the external CSM absorption is unimportant,][]{weiler02}. We therefore performed the SSA modelling of bump 1 and bump 2 using the three different values of $\beta = -1.6,~-1.3$ and $-1.0$, which correspond to ejecta density indices and deceleration parameters of $n=7$ and $m=0.8$ for radiative stars (like Wolf-Rayet [WR]), $n=12$ and $m=0.9$ for convective stars (like red supergiants), and an undecelerated blastwave expansion of $m=1$, respectively \citep{horesh13}. This range of $\beta$ is consistent with values measured from other SNe Ibc and SNe IIn \citep[see Table 1 of][]{chevalier06}.

Optical spectral observations demonstrated that SN 2014C metamorphosed into a SN IIn by the time of the second bump in the AMI light curve, which suggests that $\alpha \approx -0.7$ may be a more appropriate spectral index for the modelling of bump 2 \citep[for example see][]{weiler02,chandra09,dwarkadas16}. We therefore conducted the SSA modelling of both bumps again assuming $\alpha = -0.7$ (which implies $p=2.4$) and varying $\beta$ accordingly for $m=0.8,~0.9$ and 1. This resulted in comparable $\chi^{2}_{red}$ to those fits performed assuming $\alpha = -1$ with the best fit parameters $K_{1}$ and $K_{5}$ agreeing within an order of magnitude, but with nearly identical values of $\delta''$, for a given value of $\beta$. Since the resulting fits for both values of $\alpha$ are indistinguishable and the physical parameters derived from SSA modelling assume $p=3$ (see Section 4.2) we report the modelling results that assume $\alpha=-1$.

The radio flux density evolution subject to external FFA can be described by the parameterised model proposed by \citet{chevalier82} and adapeted by \citet{weiler86,weiler02} and \citet{vandyk94}. By altering Equation~\ref{eq:1} to include external FFA as the main absorption mechanism the model becomes:

\begin{align}\label{eq:1b}
\begin{split}
S = K_{1}~\bigg(\frac{\nu}{5~\mathrm{GHz}}\bigg)^{\alpha} \bigg(\frac{t-t_{0}}{1~\mathrm{d}}\bigg)^{\beta}~ \mathrm{e}^{-\tau_{FFA}},
\end{split}
\end{align}
where the optical depth ($\tau_{FFA}$) is given by:
\begin{align}\label{eq:2b}
\begin{split}
\tau_{FFA} = K_{2}~\bigg(\frac{\nu}{5~\mathrm{GHz}}\bigg)^{-2.1} \bigg(\frac{t-t_{0}}{1~\mathrm{d}}\bigg)^{\delta}.
\end{split}
\end{align}
In these equations $K_{2}$ is the external FFA normalisation and $\delta$ determines the time dependence of the optical depth for the FFA external absorption.

We performed the above FFA modelling of both bump 1 and bump 2 once again assuming $\alpha = -1$ and $\alpha=-0.7$. Both temporal indices can be expressed in terms of the deceleration parameter $m$ such that $\delta=-3~m$ and $\beta=-(p + 5 - 6m)/2$, which becomes $\beta=3m - 3.7$ \citep{chevalier82}. When conducting the FFA modelling we solved for $K_{1}$, $K_{2}$, and $m$. While the resulting $\chi^{2}_{red}$ were similar for all the FFA fits, the best values for $K_{1}$ and $K_{2}$ varied up to two orders of magnitude for each bump when assuming the two different spectral indices. However, the deceleration parameter $m$ agreed within the $1\sigma$ errors. As the models are indistinguishable for the two different spectral indices we assumed $\alpha = -1$ for bump 1 (while SN 2014C was classified as a SN Ib) and $\alpha=-0.7$ for bump 2 (when SN 2014C had morphed into SN IIn) when conducting the FFA modelling.

Using the above SSA and FFA equations we modelled the AMI flux measurements of bump 1 and bump 2 separately. As we only have one contiguous light curve made up of the AMI observations we set the peak frequency to $\nu_{p} = 15.7$~GHz. The modelling was conducted using the Python non-linear least-square minimization and curve-fitting package \texttt{lmfit}. The best SSA fit to each bump resulting from the three different deceleration parameters, and the corresponding reduced chi-square ($\chi^{2}_{red}$), can be found in Figure~\ref{fig:7}. We provide the best fit parameters from the SSA modelling that results in the smallest $\chi^{2}_{red}$ for bump 1 and bump 2 in Table~\ref{tab:model}, which is $m=1$ (with        $\chi^{2}_{red}=2.9$) and $m=0.8$ (with $\chi^{2}_{red}=1.6$), respectively. While we have adopted this convention it is important to note that the                        $\chi^{2}_{red}$ values are so similar that it is not possible to distinguish between the best value of $m$ for each bump. This is because all parameters in this model, particularly $K_{5}$ and $\delta''$, are heavily degenerate. However, the combination of the parameters result in a well constrained peak flux for both bumps (see discussion below). 

The results from the FFA modelling of the AMI light curve are also included in Table~\ref{tab:model}. For both bump 1 and bump 2 the $\chi^{2}_{red}$, 4.1 and 3.0, respectively, are slightly higher than those derived from SSA. However, there is still not enough of a difference in the $\chi^{2}_{red}$ to distinguish between the SSA and FFA models for each bump. We obtained a deceleration parameter of $m = 1.11 \pm 0.11$ \citep[the errors bars make this value of $m$ consistent with the requirement that $m \leq 1$][]{chevalier82} resulting in a temporal index of $\beta = -0.68 \pm 0.33$ for bump 1. For bump 2 the deceleration parameter is $m=0.63 \pm 0.07$ ($n \approx 4.7$) resulting in $\beta = -1.80 \pm 0.21$ for bump 2. The best FFA fits for bump 1 and bump 2 can be found in Figure~\ref{fig:8}.

For both the SSA and FFA modelling we present results with $m \approx 1.0$ for bump 1 and $m < 1.0$ for bump 2. We may expect that the deceleration index was close to one for bump 1 as the shock wave is young and interacting with a low density medium, and that it would reduce quite significantly when it impacts the H-shell. The values of $\beta$, $\delta$ and $\delta''$ resulting from this SSA and FFA modelling (see Table~\ref{tab:model}) are consistent with those derived for other SNe Ibc and SNe IIn \citep[e.g.][]{weiler02,chandra09,weiler11,chandra12,horesh13,dwarkadas16}.

Using the best fit parameters we optimised the models to obtain the peak flux $F_{\nu_{p}}$ and corresponding time $t_{p}$ of peak at 15.7 GHz for each bump, which are included in Table~\ref{tab:model}. These results show that the peak flux from the modelling of bump 2 was 3.7 times brighter than the peak of bump 1. We created a confidence contour map using the \texttt{lmfit} errors on each parameter to generate the $1\sigma$ error on the fit (see the shaded regions in Figure~\ref{fig:7} and \ref{fig:8}) by constraining the $\Delta \chi^{2} < 3.5$, which is appropriate for a fit involving three free parameters. This same process allowed us to obtain $1\sigma$ errors for the peak flux and corresponding time. 

The full AMI light curve and the final SSA and FFA model fits are shown in Figure~\ref{fig:1} for bump 1 and bump 2, respectively. The choice of illustrating the SSA model of bump 1 in Figure~\ref{fig:1} is because of the early time classification of SN 2014C as a SN Ib and SSA has been shown to be significant at early times for SNe Ibc \citep{chevalier06}. The FFA was chosen to be depicted for bump 2 as this is likely to be the more dominant absorption mechanism at later times, particularly for SNe with extensive CSMs \citep[e.g.][]{chandra09,chandra12}. It should be noted that it is not possible to distinguish between the SSA and FFA for the two bumps based on the goodness of fit and it is likely that more than one absorption mechanism is contributing to each density regime \citep[for example see][]{weiler02}. However, our predominantly single frequency data lacks the information we need to further constrain these effects. The corresponding measured fluxes used to model bump 1 and bump 2 are represented in Figure~\ref{fig:1} with different colours/symbols, blue diamonds and green circles, respectively. The first flux measurement included in the fitting of each bump was chosen based on when the light curve appeared to begin a steady temporal power-law brightening like a typical radio SN. This means that the early time flux measurements of SN 2014C were not used in the modelling of bump 1 and are represented by red stars in Figure~\ref{fig:1}. The observation times of each of the Keck and \citet{milisavljevic15} optical spectra are indicated by red and black triangles, respectively. 

\subsection{Radio derived physical constraints}

The time of the peak flux $F_{\nu_{p}}$ derived from the SSA modelling of the measured light curve fluxes using Equations~\ref{eq:1} and \ref{eq:2}, represents the time at which the intersection between optically thick and optically thin behaviour occurred at $\nu_{p}=15.7$ GHz \citep{chevalier98}. We can therefore use this value to estimate the blastwave radius $R$ and the magnetic field strength $B$ at the time of the intersection using Equations 13 and 14 of \citet{chevalier98}, which are repeated below for completeness. Once again assuming p = 3, the blastwave radius is calculated with the following:

\begin{align}\label{eq:3}
\begin{split}
R &= 8.8 \times 10^{15}~f_{eB}^{-1/19}~\bigg(\frac{f}{0.5}\bigg)^{-1/19}~ \bigg(\frac{F_{\nu_{p}}}{\mathrm{Jy}}\bigg)^{9/19} \bigg(\frac{D}{\mathrm{Mpc}}\bigg)^{18/19} \\
&\quad \times \bigg(\frac{\nu_{p}}{\mathrm{5~GHz}}\bigg)^{-1} \mathrm{cm}~,
\end{split}
\end{align}
with a post-shock magnetic field strength of:
\begin{align}\label{eq:4}
\begin{split}
B &= 0.58~f_{eB}^{-4/19}~\bigg(\frac{f}{0.5}\bigg)^{-4/19}~ \bigg(\frac{F_{\nu_{p}}}{\mathrm{Jy}}\bigg)^{-2/19} \bigg(\frac{D}{\mathrm{Mpc}}\bigg)^{-4/19} \\
&\quad \times \bigg(\frac{\nu_{p}}{\mathrm{5~GHz}}\bigg)~\mathrm{G}~,
\end{split}
\end{align}
 assuming equipartition $f_{eB} = 1$ (the energy density of the relativistic electrons is equal to the energy density of the magnetic fields) and that the fraction of spherical volume occupied by the emitting region (filling factor) $f=0.5$ \citep{chevalier98}. The resulting radius can be used to estimate the mean velocity of the shock at the time the source becomes optically thin at $\nu_{p}$ ($v_{s}$ = $R/t_{p}$). Both the radius and the magnetic fields resulting from this equipartition analysis then allows the derivation of the minimum internal energy of the SN radio-emitting material \citep{soderberg10}: 
 
\begin{align}\label{eq:5}
\begin{split}
E_{\mathrm{min}} = \frac{B^{2}R^{3}}{6}(1 + f_{eB})f
\end{split}
\end{align} 

As the radio emission detected from a SN explosion is directly probing the surrounding environment we can investigate the mass-loss history of the progenitor star. Using the assumption that $\rho_{\mathrm{CSM}} \propto \dot{M} / (v_{w} R^{2})$ \citep{weiler07} and the analysis from \citet{horesh13} we can calculate the electron number density of the CSM:

\begin{align}\label{eq:6}
\begin{split}
n_{e} = f_{eB} \frac{(p-2)}{(p-1)} \frac{B_{p}^{2}}{8 \pi \gamma_{m} m_{e} c^{2}} \mathrm{cm}^{-3}~, 
\end{split}
\end{align}
assuming a minimum Lorentz factor of $\gamma_{m} = 1$. 

This electron number density can then be used to estimate the progenitor's mass-loss rate prior to the explosion where:

\begin{align}\label{eq:7}
\begin{split}
\dot{M} = 4 \pi R^{2} n_{e} m_{p} v_{w}~.
\end{split}
\end{align}

Given the initial classification of SN 2014C as a stripped-envelope SN Ib, it is possible that the progenitor was a WR type star, which typically have wind speeds on the order of 1000 km s$^{-1}$. However, an optical spectrum taken by \citet{milisavljevic15}, which took place $\sim84$ days following the peak time of bump 2, predicts an unshocked wind velocity of $<100$ km s$^{-1}$. Given that the late-time optical spectra show the event has morphed into a SN IIn, this lower wind speed may be more consistent with an earlier evolution of the progenitor. We therefore calculate the mass-loss rate assuming a wind speed of 1000 km s$^{-1}$ for bump 1 and 100 km s$^{-1}$ for bump 2. The resulting radius, magnetic field, shock expansion velocity, minimum energy, electron number density and the approximate mass-loss rates of the progenitor derived from the SSA modelling of bumps 1 and 2 are presented in Table~\ref{tab:model}. The $1\sigma$ errors on these parameters are propagated through from the peak flux and time $1\sigma$ errors.

In the case that FFA is the dominant absorption mechanism we can use the deceleration parameter $m$ and the optical depth $\tau_{FFA}$ to estimate the mass loss rate of the progenitor using Equation~11 of \citet{weiler02} \citep[based on Equation~16 of][]{weiler86}:

\begin{align}\label{eq:10}
\begin{split}
\dot{M} &= 3.0 \times 10^{-6}~ \langle \tau_{\mathrm{eff}}^{0.5} \rangle~m^{-1.5}  \frac{v_{w}}{10\mathrm{~km~s}^{-1}} \bigg( \frac{v_{i}}{10^{4} \mathrm{~km~s}^{-1}} \bigg)^{1.5} \\
& \times \bigg( \frac{t_{i}}{45 \mathrm{~days}} \bigg)^{1.5} \bigg( \frac{t}{t_{i}} \bigg)^{1.5m} \bigg( \frac{T}{10^{4}~\mathrm{K}} \bigg) ~\mathrm{M}_{\odot} \mathrm{yr}^{-1}~,
\end{split}
\end{align}
where $v_{i}$ (km s$^{-1}$) is the optical expansion velocity measured at time $t_{i}$ (days), $\langle \tau_{\mathrm{eff}}^{0.5} \rangle = \tau^{0.5}_{FFA}$ for a uniform external absorbing medium and $T$ is the electron temperature in the wind \citep[usually taken to be 20,000K;][]{weiler02}. The error on the mass-loss rate is propagated through from the best fit errors on the parameter $K_{2}$. The mass loss rates derived from the FFA modelling of bump 1 and bump 2, which use the optical expansion velocity of $v_{i}=13000$ km s$^{-1}$ taken from the optical spectral observation obtained by \citet{milisavljevic15} on 2014 Jan 09 ($t_{i}=10$ days), can be found in Table~\ref{tab:model}.

\onecolumn
\begin{table}
\begin{center}
\caption{SSA and FFA model fitting and resulting parameters for bump 1 and bump 2 of the AMI light curve of SN 2014C}
\label{tab:model}
\begin{tabular}{lccccc}
\\
\hline
\hline
& \multicolumn{2}{c}{SSA fit} & & \multicolumn{2}{c}{FFA fit}\\
\cline{2-3} \cline{5-6}
\\
Parameter & Bump 1 & Bump 2 & &  Bump 1 & Bump 2\\
\hline
$\alpha$ & $-1.0^{\dagger}$ & $-1.0^{\dagger}$ & & $-1.0^{\dagger}$ & $-0.7^{\dagger}$ \\
$\beta$ & $-1.0^{\dagger}$ & $-1.6^{\dagger}$ & & $-0.68 \pm 0.33^{\mathrm{a}}$ & $-1.80 \pm 0.21^{\mathrm{a}}$ \\
\hline
$K_{1}$ & $2.20 \pm 0.16 \times 10^{3}$  & $1.57 \pm {0.34} \times 10^{6}$ & & $4.54^{+23.93}_{-3.33} \times 10^{2}$ & $7.08^{+20.07}_{-5.49} \times 10^{6}$ \\ 
$K_{2}$ & ... & ... & & $3.91^{+4.27}_{-2.59} \times 10^{6}$ & $1.01^{+1.84}_{-0.60} \times 10^{6}$ \\
$K_{5}$ & $2.76^{+22.33}_{-2.33} \times 10^{8}$ & $1.39_{-0.65}^{+1.27} \times 10^{14}$ & & ... & ... \\
$\delta$ & ... & ... & & $-3.32 \pm 0.33^{\mathrm{a}}$ & $-1.90 \pm 0.21^{\mathrm{a}}$ \\
$\delta''$ & $-3.65 \pm 0.53$ & $-4.83 \pm 0.12$ & & ... & ... \\ 
\hline
$m$ & $1.0^{\dagger}$ & $0.8^{\dagger}$ & & $1.11 \pm 0.11$ & $0.63 \pm 0.07$ \\
\textit{dof}$^{\mathrm{b}}$ & $14$ & $53$ & & 14 & 53\\
$\chi^{2}_{red}$ & 2.9 & 1.6 & & 4.1 & 3.0 \\
\hline
$t_{p}$ (days) & $78.95 \pm 3.72$ & $393.38 \pm 6.79$ & & $75.50 \pm 2.98$  & $421.21 \pm 4.97$\\
$F_{\nu_{p}}$ (mJy) & $6.63 \pm 0.30$ & $24.72 \pm 0.53$ & & $6.30 \pm 0.37$ & $23.09 \pm 0.34$ \\
\hline
$R$ ($10^{15}$ cm) & $3.32 \pm 0.07$ & $6.20 \pm 0.06$ & & ... & ... \\
$B$ (G) & $1.754 \pm 0.008$ & $1.527 \pm 0.003$ & & ... & ... \\
$v_{s}$ ($10^{3}$ km s$^{-1}$) & $4.87 \pm 0.34$ & $1.82 \pm 0.05$ & & ... & ... \\
$E_{\mathrm{min}}$ ($10^{46}$ erg) & $1.88 \pm 0.10 $ & $9.25 \pm 0.24 $ & & ... & ... \\
$n_{e}$ ($10^{4}$ cm$^{-3}$) & $7.46 \pm 0.07 $  & $5.66 \pm 0.03 $ & & ... & ... \\
$\dot{M} (10^{-4}~{\mathrm{M}}_{\odot} \mathrm{yr}^{-1})^{\mathrm{c}}$  & $2.75 \pm 0.09 $ & $0.72 \pm 0.01 $ & & $8.37 \pm 3.62$ & $5.07^{+3.44}_{-1.84}$ \\
\hline
\end{tabular}
\end{center}
All errors are 1 sigma. \\
The peak flux ($F_{\nu_{p}}$) at 15.7 GHz and the corresponding peak time ($t_{p}$) are derived from the model fit.\\
$^{~\dagger}$ Parameter was fixed during the fit.\\
$^{\mathrm{~a}}$ Derived from best fit value of $m$. See text in Section 4.1. \\
$^{\mathrm{~b}}$ degrees-of-freedom ($dof = n - p$) where $n$ = number of data-points for the given bump and $p$ = number of fitted parameters. \\
$^{\mathrm{~c}}$ Mass-loss rates calculated assuming two different progenitor wind speeds of $v_{w} =1000~\mathrm{km~s}^{-1}$ for bump 1 and $100~\mathrm{km~s}^{-1}$ for bump 2.
\end{table}
\twocolumn

\section{Discussion}

The SSA and FFA modelling described in Section 4 has been applied to the two AMI light curve bumps of SN 2014C individually. In each case the $\chi^{2}_{red}$ are similar and make it difficult to distinguish between models, although the SSA modelling may indicate a slightly better fit to both density regimes. However, these same $\chi^{2}_{red}$ are actually too high for both fits to each bump due to the low number of degrees-of-freedom. This indicates that neither of these simple models adequately describe the light curve. It is likely the light curve would be better described by models that include multiple terms for internal and external absorption \citep[see][]{weiler02} but our single frequency data prevents the fitting of these more complex models. We will therefore discuss the results and physical properties derived from the simple modelling we performed and compare them to the properties of other SNe Ibc and SNe IIn.

For the case of similar SNe Ibc, which have radio light curves that deviate from the typical power-law decline at late times \citep[i.e.][]{wellons12,corsi14}, modelling has only been applied to the earliest part of the light curve that is typical of the canonical radio temporal behaviour. Only in the case of SN 2007bg do the authors individually model the second bump of radio rebrightening \citep{salas13}. It is therefore important to note that the \citet{chevalier98} and \citet{weiler86,weiler02} formalism assumes a CSM density profile that drops off as $R^{-2}$ (see Section 4.2) and does not take into account an earlier epoch of radio brightening. Therefore any physical parameters derived for bump 2 using this modelling may not be truly representative of this second density regime. The main emphasis of the modelling of bump 2 is to investigate how well these models describe the light curve. It is also worth noting that extensive multi-frequency radio modelling of SNe IIn (the spectral classification of SN 2014C during bump 2) has demonstrated that the density profile can be much flatter with $\rho_{\mathrm{CSM}} \propto \dot{M}/(v_{w}R^{1.4-1.8})$ \citep{vandyk94,weiler02,chandra12}. The modelling of bump 1 is also likely to be overly simplistic as the radio flux density measurements preceding this bump show a deviation from the expected power-law behaviour, so there is likely to by other complex CSM structures in play. 

Contrary to results presented by \citet{wellons12} and \citet{corsi14}, which demonstrates that SSA modelling of early time radio emission of unusual radio SNe Ibc can produce sensible shock velocity and blast wave radius estimates, this same model appears to be deficient for describing the physical properties of bump 1 of SN 2014C. Bump 1 of SN 2014C peaks and becomes optically thin at 15.7 GHz around $\sim80$ days post-explosion, which is much later than most other SNe Ibc that tend to peak around 10 days near this same frequency \citep[for example see SN 1994I, SN 2004cc, SN 2004dk, SN 2004gq, and SN 2011dh,][]{alexander15,wellons12,horesh13}. However, regardless of the time of this peak, the resulting parameters such as the radius, magnetic field strength, minimum internal energy, and mass-loss rate derived from SSA modelling are consistent for both the above listed SNe Ibc and SN 2014C at the time when the radio emission becomes optically thin around 15.7 GHz. Of course the shock wave expansion velocities calculated for the same SNe Ibc are usually $>0.1c$, which is much faster than our derived expansion speed of $4870$~km~s$^{-1}$ from bump 1 of SN 2014C. This is a strong indication that either SSA is not the dominant absorption mechanism or there are other sources of absorption due to a complex CSM structure during bump 1 \citep{chevalier03}. As a result, comparing those same physical parameters at $\sim80$ days post-explosion for the above SNe Ibc show that their shock waves have expanded to $R \gtrsim10^{16}$ cm, which is well beyond our model derived radius for SN 2014C of $R=3.3 \times10^{15}$ cm at this time post-explosion. However, it is important to note that most SNe Ibc (such as those listed above) experience a power-law brightening as soon as they become detectable at radio wavelengths. This was not the case for SN 2014C, which showed early time flat or possibly erratic behaviour, similar to that seen in SN 2007bg \citep{salas13}. There are also discrepancies between the physical properties derived from our AMI radio light curve modelling results and those obtained for SN 2014C at other wavelengths as discussed below.

The early optical spectrum of SN 2014C obtained by \citet{milisavljevic15} on 2014 January 9 was reported to exhibit emission lines with velocities indicative of a 13000 km s$^{-1}$ bulk flow. This velocity is 2.7 times faster than the shock wave velocity we estimate from the SSA modelling of bump 1. However, if we assume a spherically symmetric explosion then the radio emission is expected to trace the fastest moving part of the ejecta, in this case the SN shock wave, and should be a factor of a few times faster than the bulk ejecta flow measured from optical observations. In fact, using X-ray observations \citet{margutti16pp} estimated that the low density region, which extends out to the dense H-rich shell, has a radius of $R = 5.5 \times 10^{16}$ cm. The shock wave velocity of 4870 km s$^{-1}$ predicted by our SSA modelling of bump 1 implies that this radius would not be reached for over 1000 days following the initial explosion, rather than the $\sim200$ days we know it to be from the discontinuity in the AMI radio light curve and infrared/optical/X-ray monitoring \citep{tinyanont16pp,milisavljevic15,margutti16pp}. The \citet{margutti16pp} radius is also much larger than the radius obtained from SSA modelling of bump 2 (bear in mind the caveat mentioned above), which estimates the radius at which the second radio peak becomes optically thin at 15.7 GHz (after encountering the H-shell) is a factor of 9 times smaller at $R = 6.2 \times 10^{15}$ cm. 

Conversely the velocity derived from the SSA modelling of bump 1 and bump 2 are consistent with the velocities resulting from SSA modelling of SNe IIn \citep{chevalier03}. As mentioned before, the fact that the SSA velocities derived for SN 2014C are much less than the known bulk velocity suggests that SSA is unlikely to be the only or most significant absorption mechanism. In fact, extensive studies of other SNe IIn have demonstrated that the absorption is more likely to be dominated by FFA \citep{chandra09,chandra12,dwarkadas16}.

We further speculate on other possibilities for the discrepancy between the shock velocities derived from SSA radio modelling and optical spectroscopy. One possibility could involve an asymmetric shock wave impacting a CSM shell, which would result in multiple expansion velocities. Such a scenario could involve a bipolar jet ejected along with a more slowly expanding normal SN explosion \citep[like that proposed for the unusual Type IIn SN 2010jp by][see their Figure~9]{smith12}. Evidence for ejecta asymmetry also comes from optical polarisation measurements \citep[e.g.][]{leonard06}. In the case of SN 2014C we suggest that the faster moving ejecta impacting the surrounding H-shell could be responsible for the optical emission, while a slower component could be responsible for the radio emission. Line-of-sight effects where the shock is interacting with individual dense clumps of wind material rather than a CSM shell \citep[see Figure~1a of][]{chugai94} may be another explanation for the different observed expansion velocities. In this case it is possible that the radio emission could be generated through interaction with a different and denser clump of CSM than that generating the optical emission, resulting in two different observed velocities. This scenario could also be another explanation for the double bump morphology seen in the AMI light curve where the shock wave is interacting with two CSM clumps that are at completely different locations, and therefore lines-of-sight, from the progenitor.

The SSA modelling of bump 1 estimates a mass-loss rate of $\dot{M} \approx 2.75 \times 10^{-4}~\mathrm{M}_{\odot}~\mathrm{yr}^{-1}$ during this period, which is consistent with the range of mass-loss rates seen from SNe Ibc \citep{wellons12} when assuming a wind speed of 1000 km s$^{-1}$ typical of WR stars (the main candidate for SN Ibc progenitors). This rate is at least two orders of magnitude higher than the rate estimated from the X-ray observations of SN 2014C during the low density regime \citep[][]{margutti16pp}. It is important to note that the X-ray observations and majority of optical observations used by \citet{margutti16pp} to estimate the pre-explosion and shock wave properties of the SN were not obtained during the period that AMI observed bump 1 when SN 2014C was a daytime object. The AMI flux measurements are currently the only reported observations to follow the SN shock wave during the period when it propagates through the low density region before impacting the dense H-shell, between $\sim50$ to $\sim125$ days post-explosion. 

The mass-loss rates derived from the FFA modelling are higher than those derived from SSA, particularly for bump 2. The modelling of bump 1 has a mass-loss rate of $\sim8 \times 10^{-5}- 8 \times 10^{-4}$~M$_\odot~\mathrm{yr}^{-1}$, assuming a wind velocity of $100-1000$ km~s$^{-1}$, which is at least an order of magnitude higher when compared to FFA modelling of some SN IIn \citep[for example see SN 1995N and SN 2005kd,][]{chandra09,dwarkadas16}. However, optical observations of SNe~IIn show higher mass-loss rates \citep[for example see Table~9 of][and references therein]{kiewe12} and can be of the order of $\sim0.1$~M$_\odot~\mathrm{yr}^{-1}$, producing surrounding CSM shells with masses of $\sim1$~M$_{\odot}$ within a few years pre-explosion. Such CSMs were likely caused by extreme eruptive events like those seen from LBVs \citep[for example see][]{chugai03,chugai04}. This makes the mass-loss rate of $\sim5 \times 10^{-4}$~M$_\odot~\mathrm{yr}^{-1}$ (assuming a wind speed of 100 km~s$^{-1}$) for bump 2 quite tame in comparison and is more comparable to a LBV wind rather than an actual LBV eruption \citep{smith06}. Our FFA modelling mass-loss rates are therefore not unreasonable for SNe IIn, and confirms that there were at least two distinct mass-loss episodes shortly before the SN explosion. Assuming a wind speed of 1000~km~s$^{-1}$ and 100~km~s$^{-1}$ for bump 1 and bump 2, respectively, then the FFA and SSA modelling show a modest reduction in the mass-loss rate by a factor of $1.7-3.8$ for these two density regimes. These results also demonstrate that our observation do not require a common envelope ejection scenario, which have mass-loss rate $>10^{-2}$~M$_\odot~\mathrm{yr}^{-1}$, to explain the mass-loss history of SN 2014C \citep[as suggested by][]{margutti16pp}. However, it is important to emphasise that the modelling of the AMI SN 2014C light curve presented in this paper is very simplistic given the complex environment surrounding SN 2014C but it is a good starting point for future modelling endeavours.

The AMI light curve shows a clear transition between density regimes at $\approx200$ days post-explosion indicating the time at which the SN shock began to interact with the H-shell. If we assume that the velocity of the shock was $v_{s} = 13000$~km~s$^{-1}$ then it began to interact with the H-shell at a radius of $R \approx 2 \times 10^{16}$ cm. Assuming wind velocities of $v_{w}=1000$ and $v_{w}=100$~km~s$^{-1}$, the ejection responsible for the putative H-shell, and therefore bump 2, may have occurred $\sim7$ or $\sim70$ years prior to the explosion. Both eruption dates are in good agreement with LBV stars, with the latter agreeing with the $\sim100$ year eruption date estimated by \citet{tinyanont16pp}. However, the mass-loss rates derived from the AMI light curve are much lower than other observed LBV eruptions \citep{smith06}.

\section{Conclusions}

AMI obtained 15.7 GHz radio observations of the first 567 days of the unusual stripped-envelope SN 2014C over which time it evolved from a SN Ib into a SN IIn. The resulting light curve reveals a double bump morphology that clearly indicates at least two different CSM regimes due to two distinct progenitor mass loss episodes. This behaviour has only been observed from a handful of SNe. AMI was the only reported telescope to track SN 2014C while it was a daylight object, between $\sim50$ to $\sim125$ days post-explosion. During this time we were able to observe the evolution of bump 1, which peaked at $\sim80$ days post-explosion. Variations in the light curve confirms that at an age between $\sim100-200$ days the SN shock wave began to interact with the dense H-shell proposed by \citet{milisavljevic15} and \citet{margutti16pp}, allowing us to further refine the time that SN 2014C began to metamorphous into a SN IIn. Due to this interaction the radio emission rebrightened and peaked again at $\sim400$ days, $\sim4$ times brighter than the first episode of brightening. 

Further monitoring of SN 2014C with AMI following its correlator upgrade may reveal further structure in its complex CSM as the shock wave encounters earlier episodes of mass-loss. If periodic modulations in the radio light curve like those seen in SN 2001ig and SN 2003bg \citep{ryder04,soderberg06} are observed, then this may give additional evidence of a binary interaction with SN 2014C's progenitor. \citet{ryder04} argued that the $\sim150$ days modulations seen in the radio light curve of SN 2001ig is likely due to the stellar winds from the WR progenitor colliding with the winds from a massive hot companion, resulting in the build-up of CSM on the required temporal and spatial scales. The detection of `dusty pinwheels' surrounding WR stars \citep{tuthill99}, along with the detection of the likely binary companion to SN 2001ig's progenitor \citep{ryder06}, makes such a scenario a strong possibility. Regardless, SN 2014C is not consistent with standard single massive star evolution theories that requires metallicity-dependent line-driven winds as the main mass-loss mechanism. It is highly likely that the dense shell of material impacted by SN 2014C resulted from a period of evolution involving a LBV eruption or a binary interaction as proposed by \citet{milisavljevic15}, \citet{margutti16pp}, and \citet{tinyanont16pp}. 

\section*{Acknowledgements}

Special thanks to Shrinivas Kulkarni, Yi Cao, Mansi Kasliwal and James Miller-Jones for their help and advice with this research. We thank the referee for their careful reading of the manuscript and recommendations. We also thank the staff of the Mullard Radio Astronomy Observatory for their invaluable assistance in the operation of AMI. e-MERLIN is a National Facility operated by the University of Manchester at Jodrell Bank Observatory on behalf of STFC. GEA, RPF, APR, TDS, MLP acknowledge the support of the European Research Council Advanced Grant 267697 `4 Pi Sky: Extreme Astrophysics with Revolutionary Radio Telescopes'. Research leading to these results has received funding from the EU/FP7 via ERC grant 307260; ISF, Minerva, and Weizmann-UK grants; as well as the I-Core Program of the Planning and Budgeting Committee and the Israel Science Foundation, and the Israel Space Agency. MAPT acknowledges support from the Spanish MINECO through grants AYA2012-38491-C02-02 and AYA2015-63939-C2-1-P. YCP acknowledges support from a Trinity College JRF.

\label{lastpage}

\bibliography{papers}

\end{document}